\documentclass[12pt,aps,prd,onecolumn,nofootinbib,superscriptaddress,preprintnumbers,balancelastpage]{revtex4}
\usepackage{amssymb,amsmath,latexsym,graphics, graphicx,epsfig,subfigure, multirow,comment,hyperref,appendix, slashed, verbatim} 
\usepackage{graphicx}
\usepackage{epstopdf}
\usepackage{url}
\usepackage{appendix}
\usepackage{dsfont}
\usepackage{color}
\usepackage{booktabs}
\usepackage{mathtools}
\usepackage{feynmp}
\usepackage{caption}
\newcommand{\gsim}{\lower.7ex\hbox{$\;\stackrel{\textstyle>}{\sim}\;$}}
\newcommand{\lsim}{\lower.7ex\hbox{$\;\stackrel{\textstyle<}{\sim}\;$}}
 
\def\LL{{\cal L}}

\newcommand{\TeV}{\,\text{TeV}}
\newcommand{\GeV}{\,\mathrm{GeV}}

\newcommand{\fb}{\,\mathrm{fb}}

\newcommand{\pb}{\,\mathrm{pb}}

\newcommand{\half}{{\frac{1}{2}  }}

\newcommand{\hc}{\text{ h.c. }}

\newcommand{\Br}{{\text{ Br}}}
\newcommand{\eff}{{\text{eff}}}

\begin{document}

\begin{flushright}
\mbox{\normalsize FERMILAB-PUB-13-532-T}
\end{flushright}
\vskip 80 pt

\title{Exploring the dark side of the top Yukawa} 
\author{Sonia El Hedri}
\affiliation{
SLAC, Stanford University, Menlo Park, CA 94025 USA}

\author{Patrick J. Fox}
\affiliation{Theoretical Physics Department,
Fermi National Accelerator Laboratory, Batavia, IL 60510, USA}

\author{Jay G. Wacker}
\affiliation{
SLAC, Stanford University, Menlo Park, CA 94025 USA}

\begin{abstract}
    \vskip 15 pt
    \begin{center}
    {\bf Abstract}
    \end{center}
    \vskip -8 pt
    $\quad$ 
    We investigate simple extensions of the Standard Model that could lead to the negative values of the top Yukawa coupling still allowed by the ATLAS Higgs results. Integrating out tree-level new physics generates dimension six operators that can lead to large changes to the top Yukawa couplings. If the top Yukawa coupling is negative, there is new physics beneath the $\TeV$ scale. We illustrate the simplest models still allowed by current searches.
\end{abstract}

\maketitle
\newpage

\setcounter{section}{0}
\renewcommand{\thesection}{\arabic{section}}
\renewcommand{\thesubsection}{\arabic{section}.\arabic{subsection}}
\renewcommand{\thesubsubsection}{\arabic{section}.\arabic{subsection}.\arabic{subsubsection}}
\renewcommand{\thetable}{\arabic{table}}
\makeatletter \renewcommand*{\p@subsection}{} \renewcommand*{\p@subsubsection}{}

\section{Introduction}
\label{Sec: introduction}

Since the discovery of a $125\GeV$ Higgs boson by ATLAS and CMS in July 2012 \cite{Chatrchyan:2012ufa,Aad:2012tfa}, measurements of the Higgs coupling to Standard Model particles have become a new way to probe new physics \cite{Carmi:2012yp,Englert:2011aa,Azatov:2012bz,Espinosa:2012ir,Giardino:2012ww,Li:2012ku,Rauch:2012wa,Ellis:2012rx,Klute:2012pu,Azatov:2012wq,Carmi:2012zd,Corbett:2012dm,Giardino:2012dp,Buckley:2012em,Ellis:2012hz,Montull:2012ik,Espinosa:2012im,Carmi:2012in,Plehn:2012iz,Espinosa:2012in}. Although the current measured values of these couplings are consistent with the Standard Model (SM) predictions \cite{ATLAS:2013sla,CMS:yva}, ATLAS's central value for $gg\rightarrow h \rightarrow\gamma\gamma$ \cite{ATLAS:2013oma} is significantly above the SM value, so much that there are two distinct allowed regions for the inferred top Yukawa coupling \cite{ATLAS:2013sla}. The second region is distinctive because in this region the Yukawa coupling of the Higgs to the top quark has negative values. This paper will discuss the physics necessary to accommodate this second region. Since this region is in tension with the CMS $gg\rightarrow h \rightarrow \gamma\gamma$ results, this paper will only use the ATLAS and Tevatron \cite{Aaltonen:2013ipa} results.

One possible interpretation of this new region of parameter space is the existence of a new colored and/or charged particle which runs in loops for $h\rightarrow gg$ and $h\rightarrow \gamma\gamma$.
Such a scenario has been considered in \cite{Cohen:2012wg,Dorsner:2012pp} and has been severely constrained in \cite{Reece:2012gi}. A different approach is adopted here, where minimal models which generate a negative Yukawa coupling at tree-level are studied. 

These minimal models, however, need to accommodate large changes to the top Yukawa coupling and will therefore be extremely constrained by collider searches. Integrating out tree-level physics, the lowest dimensional effective operator which could contribute to the value of the top Yukawa coupling is of the form
\begin{align}
\label{Eq: leff}
    \LL_{\eff}^{(1)} &= \frac{Y^3}{\Lambda^2} \left(H^\dagger H-\frac{v^2}{2}\right)(H\bar q_3u_3^c+h.c.)~.
\end{align}
This operator does not change the top mass and leads to tree-level contributions to the top Yukawa coupling such that
\begin{align}
y_t^{\eff} = \frac{\sqrt{2} m_t}{v} + \frac{Y^3}{\sqrt{2}}\frac{v^2}{\Lambda^2}~.
\label{Eq: effop}
\end{align}
This shows that, in order for $y_t^{\mathrm{eff}}$ to take large negative values, $\Lambda$ needs to remain close to the electroweak scale and $Y$ must be of order one. Thus, generating the effective operator shown in Eq.~\ref{Eq: leff} then requires introducing new sub-$\TeV$ particles.

\begin{figure}[!h]
    \centering
    \begin{fmffile}{tpartner2}
        \begin{fmfgraph*}(120,100)
            \fmfset{arrow_len}{2mm}
            \fmfset{arrow_ang}{20}
            \fmfset{wiggly_len}{5mm}
            \fmfset{wiggly_slope}{70}
            \fmfleft{i1,i2}
            \fmftop{o1}
            \fmfright{o2,o3}
            \fmf{dashes,label=$h$}{i2,v1}
            \fmf{fermion,tension=1.2,label=$q_3$,label.side=left}{i1,v1}
            \fmf{fermion,label=$U^c$,label.side=right}{v1,v2}
            \fmf{dashes,tension=0.1,label=$h$}{v2,o1}
            \fmf{fermion,label=$Q$}{v2,v3}
            \fmf{fermion,tension=1.2,label=$u^c$,label.side=left}{o2,v3}
            \fmf{dashes,label=$h$}{v3,o3}
        \end{fmfgraph*}
    \end{fmffile}
    \quad\quad
    \begin{fmffile}{tpartner1}
        \begin{fmfgraph*}(120,100)
            \fmfcmd{%
                vardef cross_bar (expr p, len, ang) =
                ((-len/2,0)--(len/2,0))
                rotated (ang + angle direction length(p)/2 of p)
                shifted point length(p)/2 of p
                enddef;
                style_def crossed expr p =
                cdraw p;
                ccutdraw cross_bar (p, 5mm, 45);
                ccutdraw cross_bar (p, 5mm, -45)
                enddef;
            }
            \fmfset{arrow_len}{2mm}
            \fmfset{arrow_ang}{20}
            \fmfset{wiggly_len}{5mm}
            \fmfset{wiggly_slope}{70}
            \fmfleft{i1,i2}
            \fmfright{o1,o2,o3}
            \fmf{fermion,tension=1.2,label=$q_3$}{i1,v1}
            \fmf{dashes,label=$h$}{i2,v1}
            \fmf{fermion,label=$U^c$}{v1,v2}
            \fmf{fermion,label=$U$,label.side=left}{v3,v2}
            \fmf{crossed,tension=0}{v1,v3}
            \fmf{fermion,tension=1.2,label=$u^c$,label.side=left}{o1,v3}
            \fmf{phantom}{v3,o3}
            \fmffreeze
            \fmf{dashes,label=$S$,label.side=left}{v3,v4}
            \fmf{dashes,tension=2.5}{o2,v4}
            \fmf{dashes,tension=2.7,label=$h$}{v4,o3}
            \fmflabel{$h$}{o2}
        \end{fmfgraph*}
    \end{fmffile}
    \quad\quad\quad
    \begin{fmffile}{2hdm}
        \fmfframe(1,7)(1,7){
        \begin{fmfgraph*}(120,100)
            \fmfset{arrow_len}{2mm}
            \fmfset{arrow_ang}{20}
            \fmfset{wiggly_len}{5mm}
            \fmfset{wiggly_slope}{70}
            \fmfleft{i1,i2,i3}
            \fmfright{o1,o2}
            \fmf{dashes,label=$h$,label.side=left}{i1,v1}
            \fmf{dashes,tension=0}{i2,v1}
            \fmf{dashes,label=$h$,label.side=left}{i3,v1}
            \fmf{dashes,label=$\Phi$}{v1,v2}
            \fmf{fermion,label=$u^c$}{o1,v2}
            \fmf{fermion,label=$q_3$,label.side=left}{o2,v2}
            \fmflabel{$h$}{i2}
        \end{fmfgraph*}
    }
    \end{fmffile}
    \caption{Possible tree level Feynman diagrams associated to the dimension 6 operator in Eq.~\ref{Eq: leff}. On the left, the new particles contributing to the top Yukawa are two top quark partners $Q$ and $U$ (model 1), in the middle, one top quark partner $U$ and one singlet $S$ (model 2), and on the right, a new electroweakly charged scalar $\Phi$ (model 3).\label{Fig: effdiagrams}}
\end{figure}
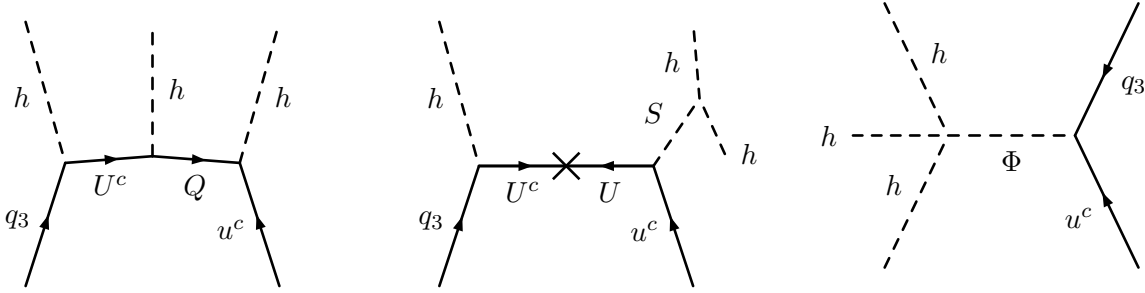

This dimension six operator arises in three simple scenarios, where the new particles can be
\begin{enumerate}
    \item new vector-like fermions: two Dirac top partners, $Q$ and $U$ (Fig.~\ref{Fig: effdiagrams}-a)
    \item both fermions and scalars: one fermionic top quark partner $U$ and one scalar singlet $S$ (Fig.~\ref{Fig: effdiagrams}-b)
    \item new electroweakly charged scalars (Fig.~\ref{Fig: effdiagrams}-c).
\end{enumerate}
which lead to the diagrams shown in Fig.~\ref{Fig: effdiagrams}. In order to maximize the effective operator contribution to the top Yukawa, these particles would have to be light and have large couplings to the Higgs boson and the top. Therefore, negative top Yukawas require significant changes to sub-$\TeV$ physics that have been searched for by Tevatron and LHC searches. Due to the large values of the couplings of the new particles to the SM, models with negative top Yukawas also lead in most cases to low scale Landau poles.

We discuss below the regions still allowed by the Higgs searches at ATLAS, before going on to describe possible models to realize the negative top Yukawa solutions, the so called ``dark side".
In Section \ref{Sec: Top Partners} we discuss models that introduce new vector-like top quark partners.  In Section \ref{Sec: h2} we describe models with extra Higgs fields.  Specifically, we consider a two Higgs doublet model \cite{Lee:1973iz,Branco:2011iw,Gunion:2002zf} where both Higgses couple to the top quark, an arbitrary number of Higgs doublets, and finally a model involving higher $SU(2)$ representations.  We show that in all but the last case these models cannot generate a sufficiently negative top Yukawa to reach the dark side.  The last model discussed can and has many new light states that can be searched for at the LHC.


\subsection{ATLAS searches}
\label{Sec: searches}
\begin{figure}
\centering
\includegraphics[width=0.5\linewidth]{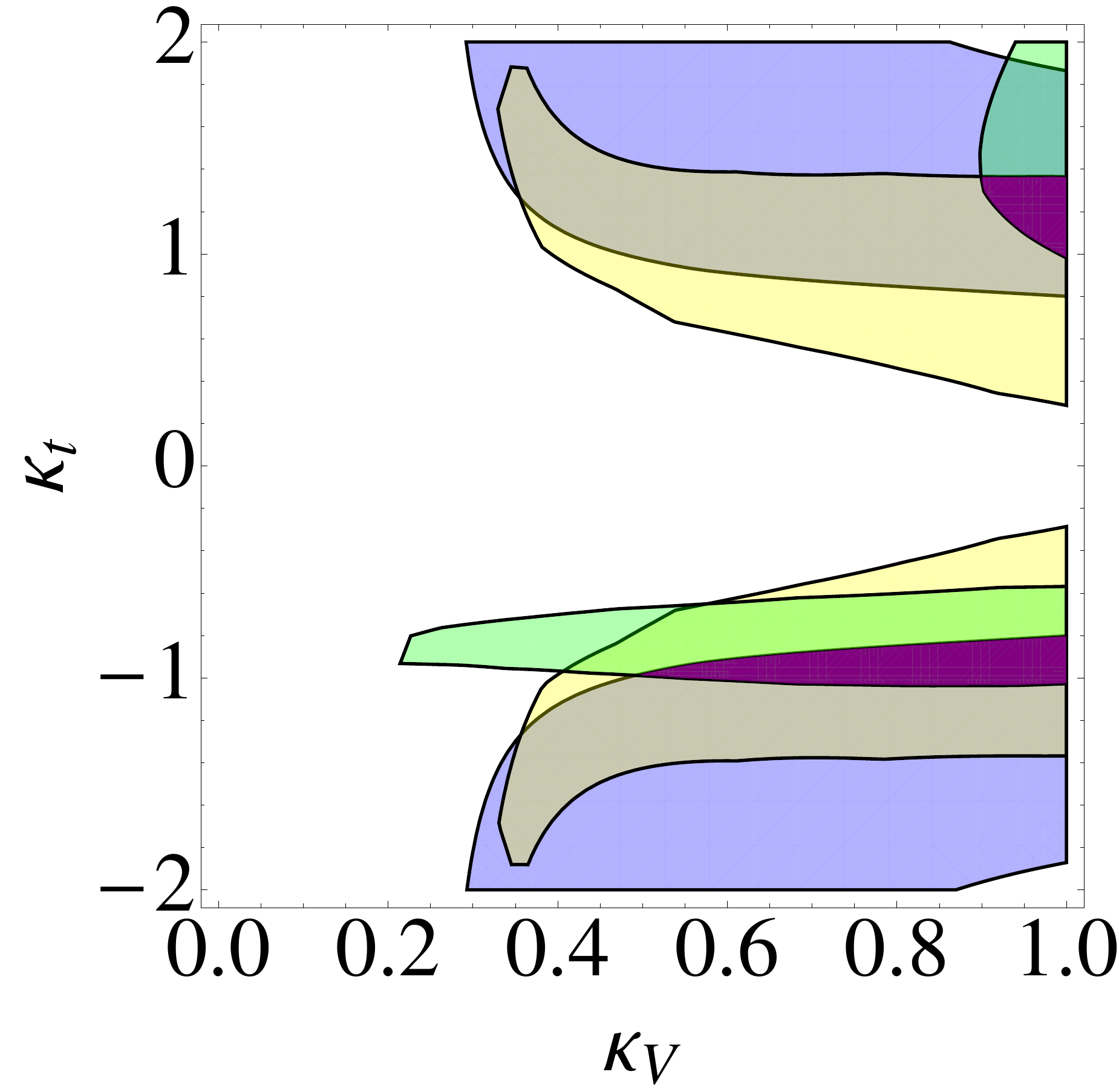}
\caption{\label{Fig: boundsATLAS} $2\sigma$ bounds on $\kappa_t$ and $\kappa_V$ from the ATLAS $\rightarrow\gamma\gamma$ (green), $h\rightarrow WW\rightarrow 2l2\nu$ (yellow) and $h\rightarrow ZZ\rightarrow 4l$ (blue) searches. The intersection of all these bounds is in purple.}
\end{figure}

Throughout this paper, there are two important Higgs couplings that will be modified with respect to their SM values: the Higgs couplings to the top quark and to vector bosons. Their deviations with respect to their SM values, $\kappa_t$ and $\kappa_V$, are defined by
\begin{align}
\mathcal{L} \supset \frac{m_t}{\sqrt{2} v}\kappa_t h t\bar t + \kappa_V \frac{g}{\sqrt{2}v}\left(m_W^2 W^2+ \frac{1}{2}m_Z^2 Z^2\right)h~.
\label{Eq: dev}
\end{align}
In the SM, 
$\kappa_t = \kappa_V = 1$.
The $2\sigma$ bounds on these parameters from the ATLAS $h\rightarrow\gamma\gamma$ \cite{ATLAS:2013oma}, $h\rightarrow WW\rightarrow 2l2\nu$ \cite{ATLAS:2013wla} and $h\rightarrow ZZ\rightarrow 4l$  \cite{ATLAS:2013nma} are shown in Fig.~\ref{Fig: boundsATLAS}. The allowed regions in the  ($\kappa_t$, $\kappa_V$) space are defined by the intersection of the three bounds mentioned above and are therefore larger than the regions we would find using a full statistical analysis.  As can be seen from Fig.~\ref{Fig: boundsATLAS} the non-SM region of top Yukawa still allowed by the data is $-1\lsim \kappa_t\lsim -0.8$.

\section{Fermionic top partners}
\label{Sec: Top Partners}

The effective operator shown in Eq.~\ref{Eq: leff} can be generated by introducing fermionic top partners. The diagrams in Fig.~\ref{Fig: effdiagrams} show two classes of minimal models. The corresponding new particles could be either a pair of top partners $Q$ and $U$, or one top partner $Q$ or $U$ and a scalar $\eta$ with charge $(1,0)$ or $(3,0)$ under $SU(2)\times U(1)$. Here, we focus on extensions of the SM with either two top partners or one top partner and a scalar singlet.

Models studied in this section involve light Dirac top partners and are therefore strongly constrained by the corresponding ATLAS \cite{ATLAS:2013di}  and CMS \cite{CMS:2013ima} searches.  The top partner production cross section is dominated by pair production through gluons and is therefore model independent. Branching ratios associated to the three top partner decay modes $Q, U\rightarrow W b, tH$ and $tZ$ depend however on the $SU(2)$ quantum numbers and Yukawa couplings of the top partners. Both CMS and ATLAS searches accommodate all the corresponding possibilities by giving bounds on the top partner mass for all possible combinations of the three branching ratios. Throughout this section, we use the loosest possible bound on the top quark mass, which is around $650\GeV$.
\subsection{Two top partners}
\label{Sec: twotop}

The effective operator shown in Eq.~\ref{Eq: leff} can be generated by extending the SM with two vector-like fermions $(Q,Q^c)$ and $(U,U^c)$ whose quantum numbers are:
\begin{align}
    Q \sim\; (3, 2, \dfrac{1}{6})\quad\quad Q^c\; (\bar 3, 2, -\dfrac{1}{6})\quad\quad U \sim \;(3, 1, \dfrac{2}{3})\quad\quad U^c \; (\bar 3, 1, -\dfrac{2}{3})~.
\end{align}
Similar fields can be found in Little Higgs \cite{ArkaniHamed:2002qy,Han:2003wu,Schmaltz:2005ky} and Randall-Sundrum \cite{Randall:1999ee} models. The corresponding Lagrangian is
\begin{align}
    \LL_{\mathrm{int}} = y_0 hq_3u^c + y_1 hq_3 U^c - y_2 h Q U^c + y_3 h Q u^c + M_1 U U^c + M_2 Q Q^c
    \label{Eq: ltop}
\end{align}
Since this model involves two new fields, only two of the three new Yukawa couplings -- $y_1$, $y_2$ and $y_3$ -- can be made positive by phase redefinition. Here, we define $y_0$, $y_1$ and $y_3$ to be positive while the sign of $y_2$ remains free.  With the parameterization of Eq.~\ref{Eq: ltop}, in order to get a negative Higgs coupling to the top quark mass eigenstate, $y_2$ will then have to be positive.
The corresponding mass matrix is
\begin{align}
    M = \begin{pmatrix}
        \dfrac{y_0 v}{\sqrt{2}} & \dfrac{y_1 v}{\sqrt{2}} & 0\\
        0 & M_1 & 0\\
        \dfrac{y_3 v}{\sqrt{2}} & \dfrac{-y_2 v}{\sqrt{2}} & M_2
    \end{pmatrix}
\end{align}
This new model involves three states mixing with each other and has six new parameters. However, viable models need to obey the following four constraints: 
\begin{itemize}
    \item the lightest mass eigenstate has mass $m_t$
    \item the other top quark partners have to be heavier than the lower mass bounds set by the ATLAS  \cite{ATLAS:2013di}   and CMS \cite{CMS:2013ima} searches. 
    \item the Yukawa couplings have to remain perturbative up to about $10\TeV$
    \item $-1<\kappa_t<-0.8$ where $\kappa_t$ is the ratio of the top Yukawa coupling over its SM value.
\end{itemize}
For small mixing angles, satisfying this last constraint requires large Yukawa couplings, which are likely to lead to low scale Landau poles.  As we will see, even if we ignore the second constraint there is no way to satisfy the other requirements.

\begin{figure}
\centering
\includegraphics[width=0.6\linewidth]{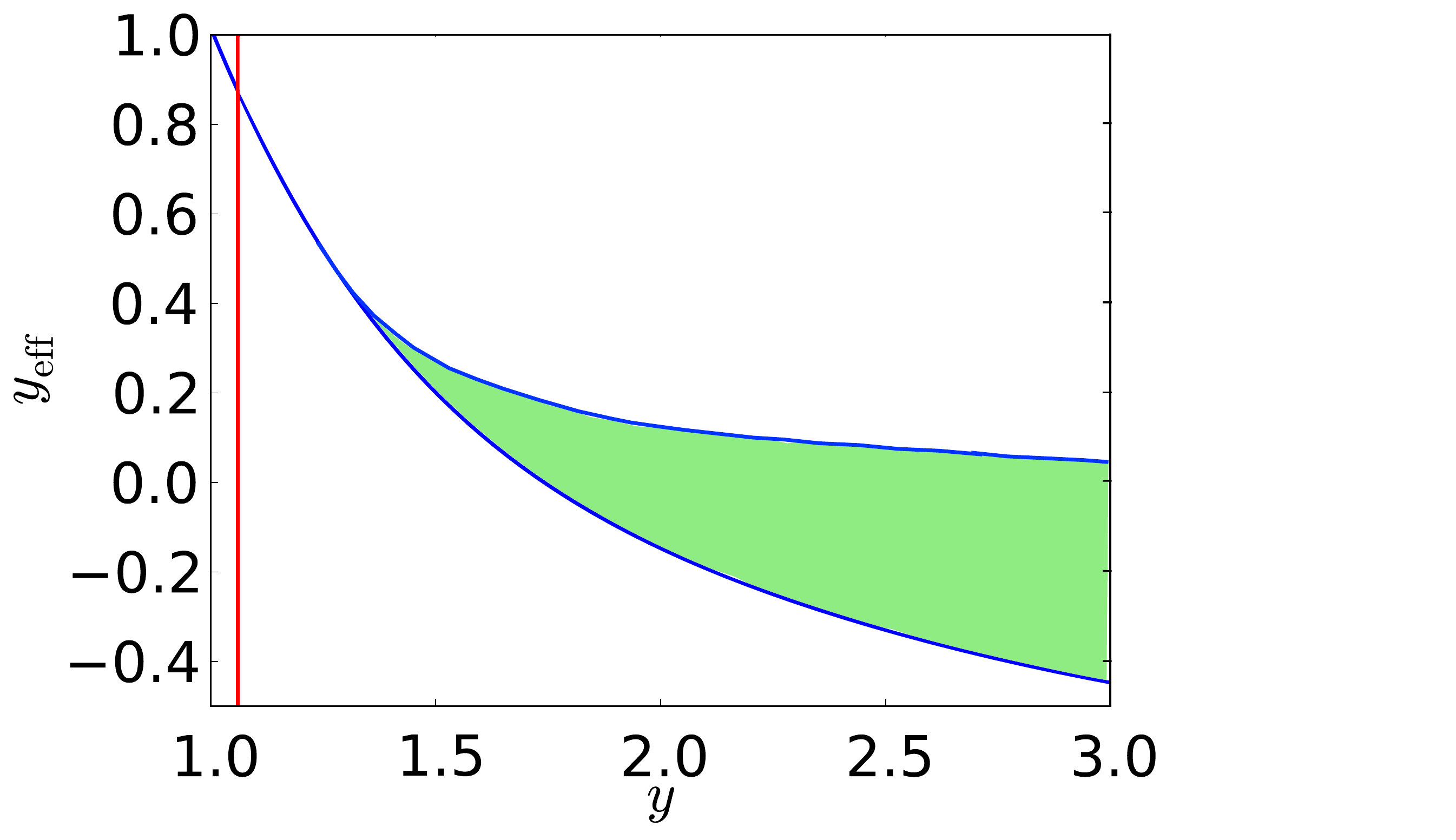}
\caption{\label{Fig: scan3}Range of $y_{\mathrm{eff}}$ vs $y$ for $y = y_0 = y_1 = y_2 = y_3 $ and $M_1,M_2>100\GeV$ (green).The red line shows the maximal value of $y$ for which there are no Landau poles below $10\TeV$.}
\end{figure}

The RGEs for the top Yukawa couplings \cite{Machacek:1983fi} are shown in Appendix \ref{Sec: RGE2top}. When all the Yukawa couplings are equal, the system exhibits a $\mathds{Z}_2\times  \mathds{Z}_2$ symmetry which allows to show that the configuration for which the perturbativity bounds are the loosest is the one with 
\begin{align}
|y_0| = |y_1| = |y_2| = |y_3| = y~,
\label{Eq: yeq}
\end{align}
for which Landau poles below $10\TeV$ are avoided if and only if
$y \lsim 1.06$.
Fig.~\ref{Fig: scan3} shows the range of possible values for $y_{\mathrm{eff}}$ in function of $y$ for positive $y$ and 
$M_1,M_2 > 100\GeV$.
The minimum $y_{\mathrm{eff}}$ that can be reached here is 
\begin{align}
y_{\mathrm{eff}}\sim 0.9~.
\end{align}
Models not satisfying Eq.~\ref{Eq: yeq} will in general be more constrained by the perturbativity requirement and are therefore not expected to give significantly better results than the simple model outlined here. 
To show this, we expand away from the $\mathds{Z}_2$ limit by scanning over the full $5$D parameter space. As in the previous section, the Yukawa coupling RGEs (shown in Appendix \ref{Sec: RGE2top}) forbid Yukawa couplings greater than $1.06$. 

We scan over  $[-2, 0]$ for $y_2$ and $[0,2]$ for $y_1$ and $y_3$ with a step size of $0.05$. $M_1$ and $M_2$ are scanned over the range $[400\GeV, 1000\GeV]$ with a step size of $100\GeV$. $y_0$ is fixed by requiring the lightest mass eigenstate to be the top quark. Requiring the heavy top quarks to be heavier than $650\GeV$ and the Yukawa couplings to remain perturbative up to $10\TeV$, the minimum $y_{\mathrm{eff}}$ that can be reached is
is
\begin{align}
y_{\mathrm{eff}} \sim 0.73~.
\end{align}
Since the top Yukawa coupling $y_{\mathrm{eff}}$ is a well-behaved function of the different parameters scanned over and the step size is small, the existence of regions of parameter space with a Yukawa coupling close to minus one and no low scale Landau poles is highly unlikely. 

\subsection{Top partner plus singlet}
\label{Sec: topsinglet}

Extending the SM with a vector-like top quark partner $Q$ ($SU(2)$ doublet) or $U$ ($SU(2)$ singlet) and a scalar singlet $S$ leads to the second diagram of Fig.~\ref{Fig: effdiagrams}. The Lagrangians for these simple models are
\begin{align}
    \LL_{int} &= y_0 hq_3u^c + y_1 SU u^c + y_2 Hq_3U^c + \lambda' Sh^\dagger h + M UU^c+\half M_{S}^2 S^2\\
    \LL_{int} &= y_0 hq_3u^c + y_1 Sq_3 Q^c + y_2 HQu^c + \lambda' Sh^\dagger h + M QQ^c+\half M_{S}^2 S^2
\end{align}
where $q$, $u^c$ and $H$ are the interaction eigenstates corresponding to the left and right-handed top quarks and the Higgs doublet respectively. 
In what follows, we also choose $S$ to be a complex scalar. This choice affects only the RGEs, which would lead to tighter bounds on the values of the Yukawa couplings if $S$ is a real scalar. If the $S$ does not get a vev, the fermion mass matrix is
\begin{align}
    M_F &=  \begin{pmatrix}
        \dfrac{y_0v}{\sqrt{2}} & \dfrac{y_2 v}{\sqrt{2}}\\
        0 & M
    \end{pmatrix}
\end{align}
The mass of the heavy top partner can then be expressed as
\begin{align}
    m_T &= m_t\dfrac{y_0}{y_{SM}}\sqrt{1+\dfrac{y_2^2}{y_0^2 - y_{SM}^2}}
    \label{Eq: mT}
\end{align}
The left and right-handed mass eigenstates are characterised by the mixing angles $\beta$ and $\alpha$ respectively, given by
\begin{align}
    \sin\alpha &= \dfrac{1}{\sqrt{1+z^2 (x^2-1)}},\quad\sin\beta = \dfrac{x}{\sqrt{1+z^2(x^2-1)}}~,
\end{align}
where
\begin{align}
    x &= \dfrac{y_{SM}}{y_0}\dfrac{m_T}{m_t} > 1 \text{, }\quad z = \dfrac{y_0}{\sqrt{y_0^2-y_{SM}^2}} > 1~.
\end{align}
EWSB induces a mixing between $h^0$ and $S$, with mixing angle $\theta$. Diagonalizing the scalar mass matrix gives two mass eigenstates: $h_1^0$ (the $125\GeV$ Higgs) and $h_2^0$, defined by
\begin{align}
    h_1^0 &= \cos\theta h^0 - \sin\theta S\quad h_2^0 = \sin\theta h^0 + \cos\theta S.
\end{align}
The coupling of $h$ to the top quark mass eigenstate, $y_{\mathrm{eff}}$, is then given by
\begin{align}
    y_{\mathrm{\eff}} &= y_0 \cos\theta \cos\beta\cos\alpha + y_1 \sin\theta \sin\beta\cos\alpha + y_2 \cos\theta \cos\beta\sin\alpha
    \label{Eq: angles}
\end{align}
Using Eqs \ref{Eq: angles} and expressing $\alpha$ and $\beta$ in function of $y$, $y_1$ and $m_T$, we get 
\begin{align}
    |y_1| = \dfrac{(y_{\mathrm{SM}}^2 m_T^2 - y_0^2 m_t^2)(2y_0^2 - y_{\mathrm{SM}}^2)\cos\theta - \kappa_t y_{\mathrm{SM}}^2 y_0^2 (m_T^2 - m_t^2)}{m_T y_0^2\sin\theta\sqrt{(y_0^2 - y_{\mathrm{SM}}^2)\left(\frac{m_T^2}{y_0^2} - \frac{m_t^2}{y_{\mathrm{SM}}^2}\right)}}
    \label{Eq: y1scale}
\end{align}

\begin{figure}
\centering
\includegraphics[width=0.48\linewidth]{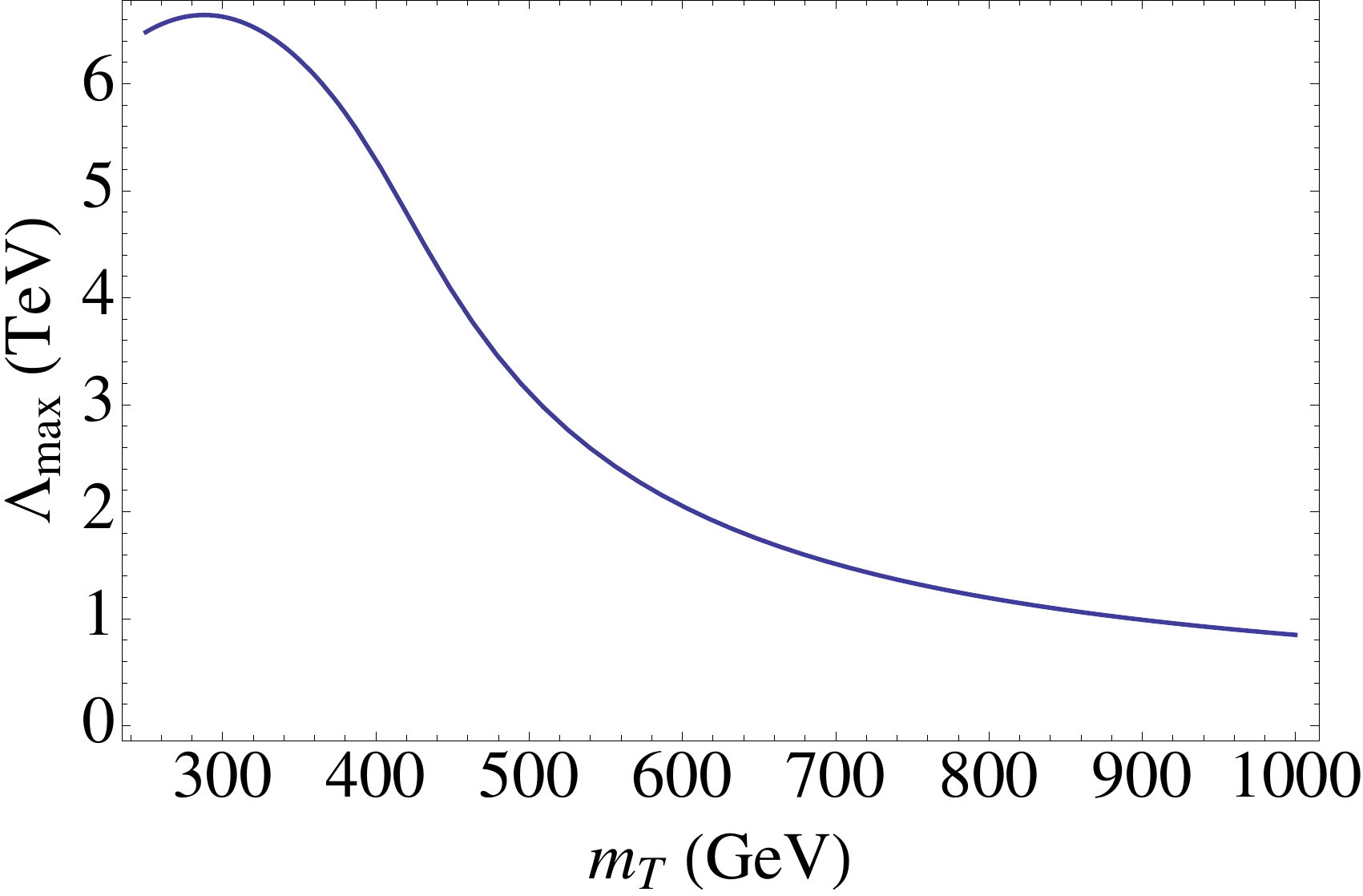}
\includegraphics[width=0.5\linewidth]{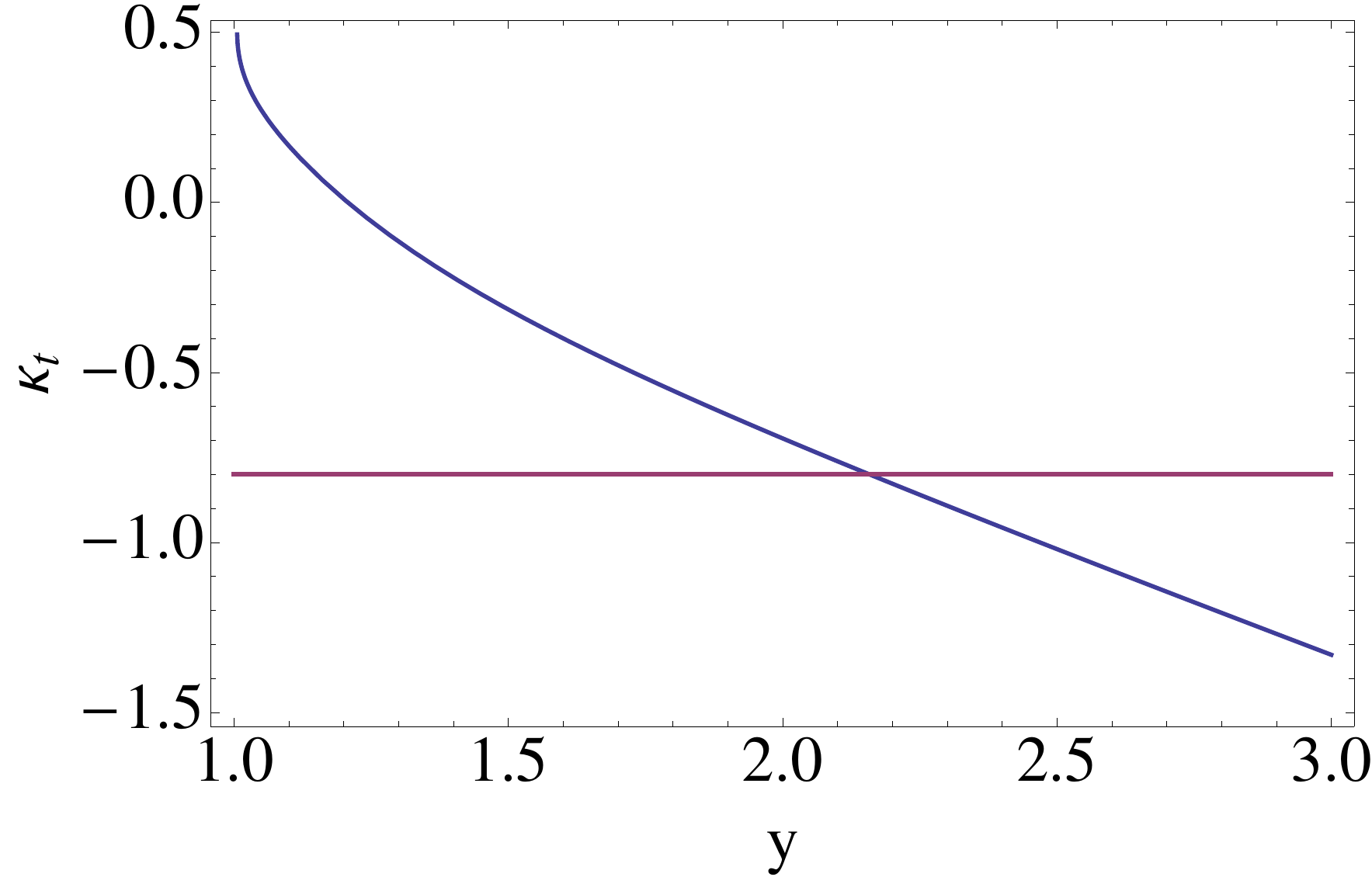}
\caption{\label{Fig: topsinglet} Maximum possible scale for the Landau poles (in $\TeV$) in function of $m_T$ ($\GeV$) for $\kappa_t = -0.8$ and $\kappa_V = 0.5$ (left). $\kappa_t$ in function of $y = y_0 = y_1 = y_2$ (right).}
\end{figure}

The limits on $\kappa_t$, $\kappa_V$ from the ATLAS Higgs searches \cite{ATLAS:2013sla} for this model are shown in Fig.~\ref{Fig: boundsATLAS}. For negative $\kappa_t$, the allowed 
region is
    $\kappa_V \ge 0.5$ and $-1 \le \kappa_t \le -0.8$.
The case of
 $\kappa_V = 0.5$ and $\kappa_t = -0.8$
yields the lowest possible values of $|y_1|$ for given $y_0$, $m_T$. For this choice of parameters and using the RGEs shown in App.~\ref{Sec: RGE1top}, Fig.~\ref{Fig: topsinglet} shows the maximum possible energy scales $\Lambda$ that can be reached in function of $m_T$. As shown in Sec.~\ref{Sec: searches}, the lowest possible bound on $m_T$ from the ATLAS  and CMS searches \cite{ATLAS:2013ima,CMS:2013ima} is around $650\GeV$, which would correspond to Landau poles below $2\TeV$. Models where the top Yukawa coupling is driven negative by introducing one Dirac top partner and one scalar singlet can therefore be excluded.

\section{Introducing additional scalars}
\label{Sec: h2}

The effective operator shown in Eq.~\ref{Eq: effop} can be generated by introducing new scalar fields as shown in the rightmost diagram of Fig.~\ref{Fig: effdiagrams}. Since these scalars have to couple both to the Higgs and to the top quarks, they have to be $SU(2)$ doublets. Since the corresponding models do not involve light new colored particles, they are expected to be less constrained than the models involving top partners studied in the previous section. A minimal scenario which could lead to a negative top Yukawa coupling would then be a two Higgs doublet model  \cite{Lee:1973iz,Branco:2011iw,Gunion:2002zf}  in which the second Higgs interaction eigenstate has a large negative Yukawa coupling to the top quark.

\subsection{A minimal model}
\label{Sec: minimal}

The minimal Lagrangian needed to obtain the operator in Eq.~\ref{Eq: effop} is 
\begin{align}
    \LL = (y h_1+ y' h_2) q_3 u^c_3 - \lambda ( |h_1|^2 - v^2/2)^2 - \frac{1}{2}m_{h_2}^2|h_2|^2 - \lambda' h_2^{\dagger} h_1( |h_1|^2 - v^2/2) +\hc
    \label{Eq: lagrangian}
\end{align}
In this preliminary study, many terms usually encountered in two Higgs doublet models are set to zero for simplicity. We show later that introducing them does not modify our conclusions. One benefit of this simple model is that the new interaction eigenstate $h_2$ does not get a vev. The interaction eigenstates $h_1$ and $h_2$ are then of the form
\begin{align}
    h_1 = \begin{pmatrix}
        G^\pm\\
        \dfrac{v + h_1^0 + i G^0}{\sqrt{2}}
    \end{pmatrix}
    \quad
    h_2 = \begin{pmatrix}
        H^\pm\\
        \dfrac{h_2^0 + i A^0}{\sqrt{2}}
    \end{pmatrix}
\end{align}
The $\lambda'$ quartic term induces a mixing between $h_1^0$ and $h_2^0$ with a mixing angle $\theta$. In what follows, we name the $125\GeV$ mass eigenstate $h^0$ and the other mass eigenstate $H^0$. Since the concept of a negative top Yukawa coupling is ambiguous if $h^0$ and $H^0$ are indistinguishable, we consider only the case where the mass difference between the two Higgses is greater than $3\GeV$, the mass resolution of ATLAS and CMS.

\begin{figure}
 \centering
 \includegraphics[width=0.5\linewidth]{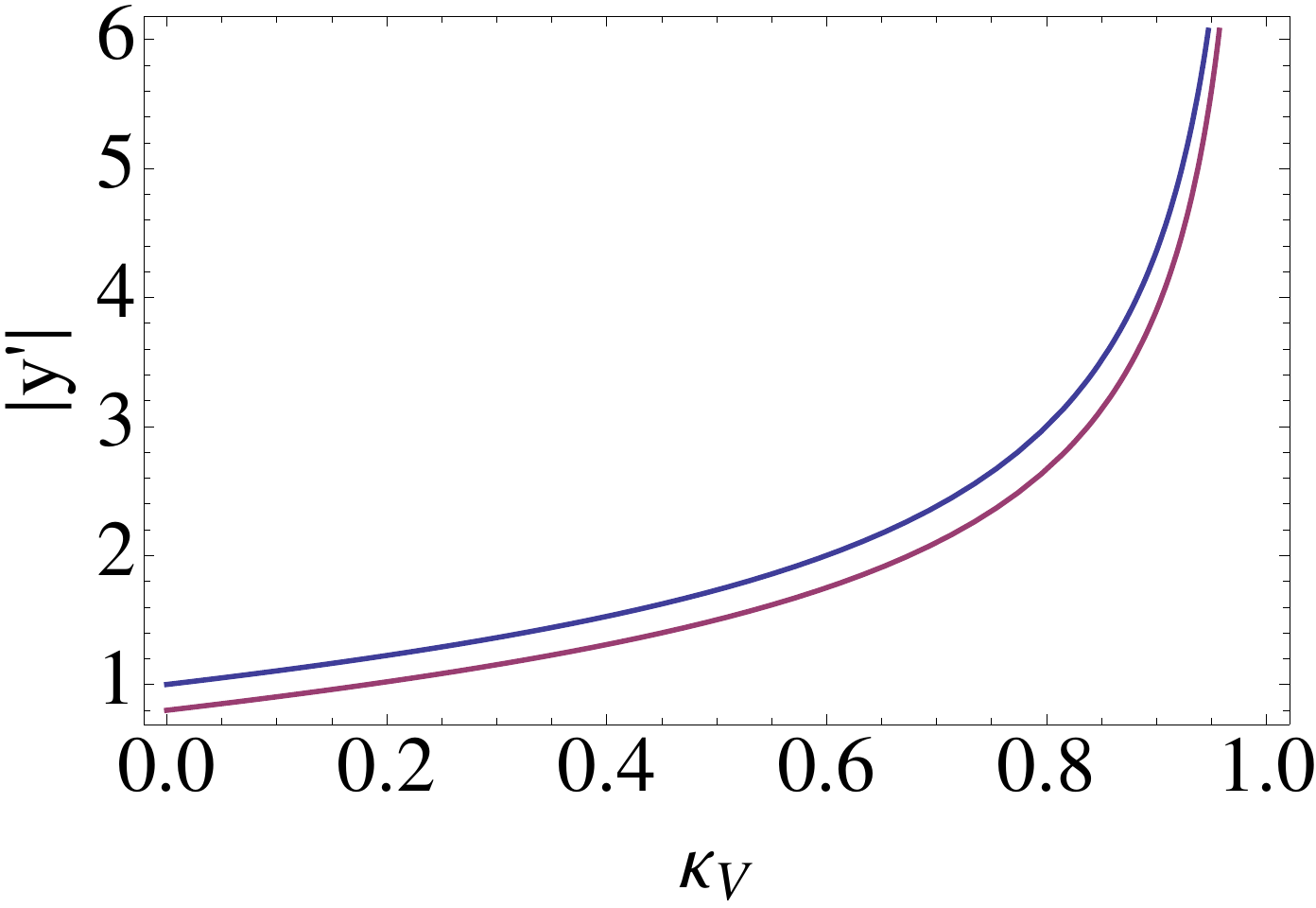}
 \caption{\label{Fig: ypkv}$|y'|$ in function of $\kappa_V$ for $\kappa_t = -1$ (blue) and $\kappa_t = -0.8$ (red)}
 \end{figure}

The interaction Lagrangian for the $125\GeV$ Higgs mass eigenstate $h^0$ is
\begin{align}
\nonumber
\LL &=  \frac{m_t \cos\theta}{\sqrt{2}v} ( 1 + \tan\theta r) h^0t t^c+(m_W^2 W^2 + m_Z^2 Z^2/2)h^0 \cos \theta
\end{align}
with $r = y'/y$. 
The couplings $y$ and $y'$ can be rewritten in function of $\kappa_t$ and $\kappa_V$ to give
\begin{align}
y& = y_{SM}\quad y'= y_{SM}\dfrac{\kappa_t-\kappa_V}{\sqrt{1-\kappa_V^2}}
\label{Eq: yp}
\end{align}
Fig.~\ref{Fig: ypkv} shows the dependance of $y'$ on $\kappa_V$ for $\kappa_t = -1, -0.8$.From this, we see that for 
$\kappa_V\gsim 0.5$
the Yukawa couplings are likely to become non perturbative at the $\TeV$ scale. 
 Keeping $y'$ low will then require a large mixing angle between the two Higgses, leading to a significant suppression of the Higgs detection rates for vector boson fusion and associated production. This minimal two-Higgs model will then be strongly constrained by the current Higgs results.
 
\begin{figure}
\centering
\includegraphics[width=0.45\linewidth]{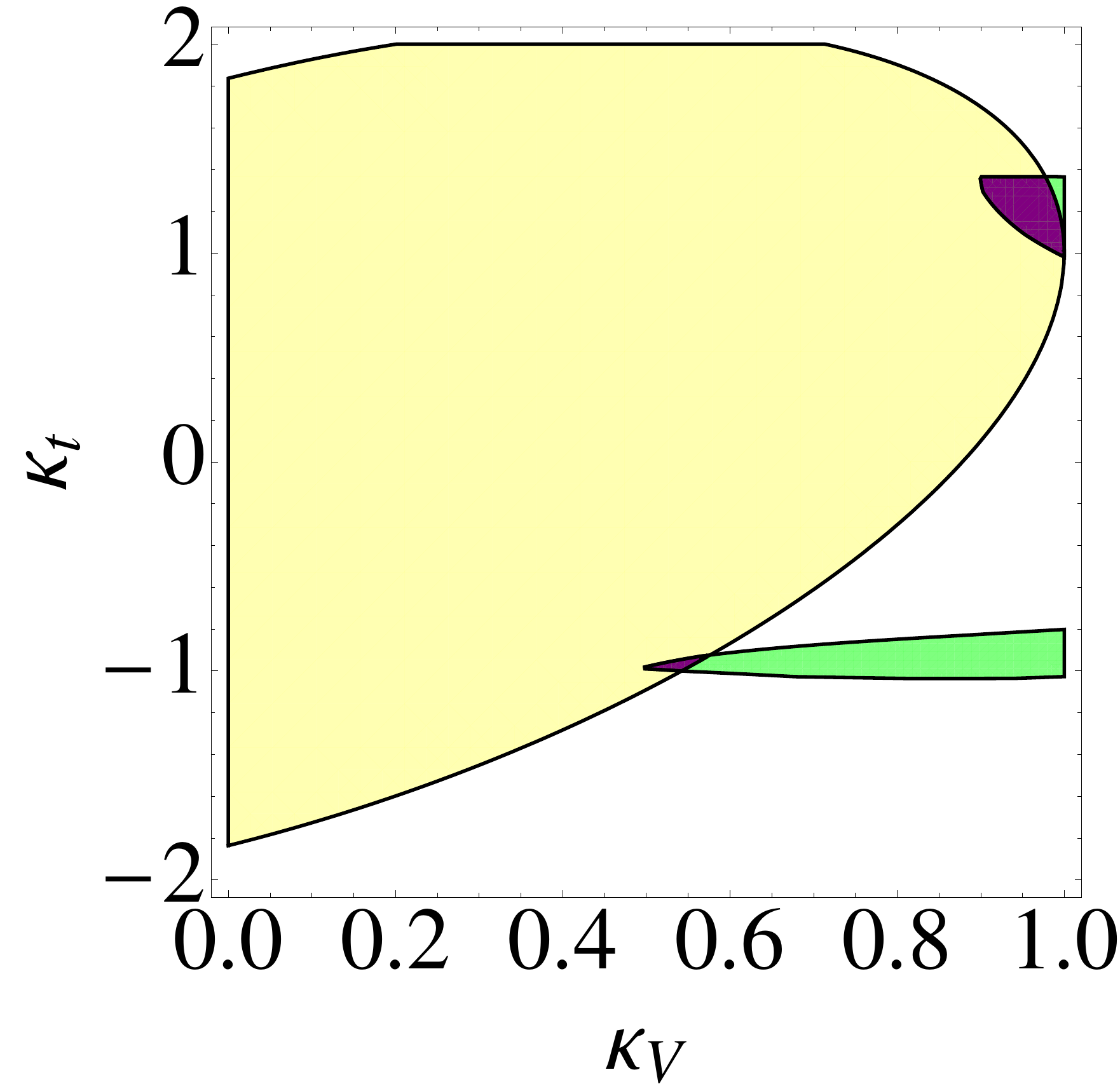}\includegraphics[width=0.45\linewidth]{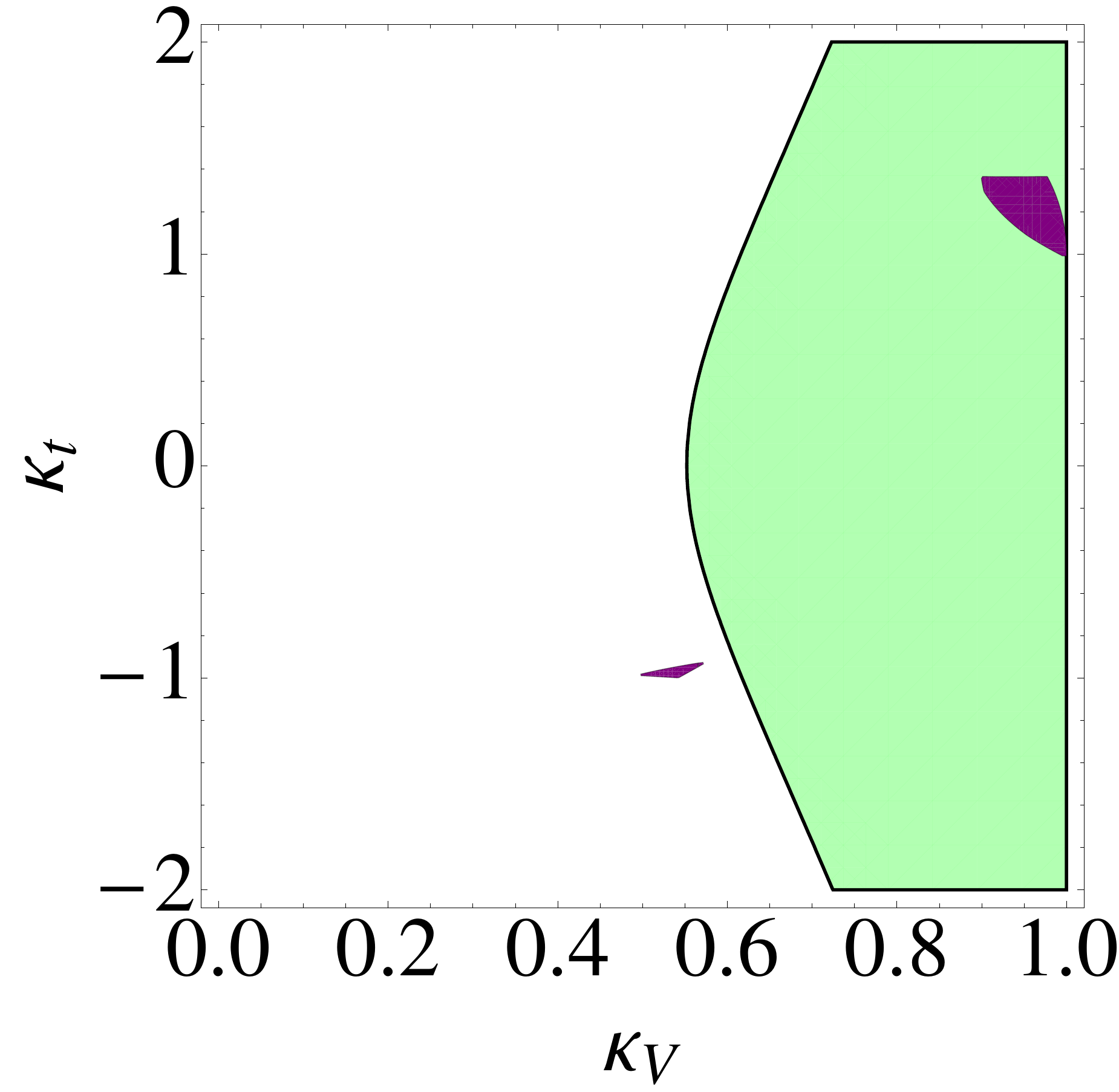}
\caption{Left: Values of $\kappa_t$, $\kappa_V$ for which the top Yukawa couplings become perturbative only above $10\TeV$ (yellow) and intersection of the regions allowed by the ATLAS $h\rightarrow \gamma\gamma$ \cite{ATLAS:2013oma}, $h\rightarrow WW\rightarrow 2l2\nu$ \cite{ATLAS:2013wla} and $ZZ\rightarrow 4l$ \cite{ATLAS:2013nma} searches (light green). The intersection of all the regions is shown in purple. Right: Intersection of the ATLAS and perturbativity bounds (purple) and region allowed by the combined CDF and D0 analysis for $WH\rightarrow b\bar b$ \cite{Aaltonen:2013ipa} (green).\label{Fig: RGE}}. 
\end{figure}

The RGEs for $y$ and $y'$ are
\begin{align}
16\pi^2\frac{dy}{d\log\mu} &= \frac{9}{2}\left(y^2+y'^2\right)y - 8 y g_s^2\\
16\pi^2\frac{dy'}{d\log\mu} &= \frac{9}{2}\left(y^2+y'^2\right)y' - 8 y' g_s^2
\label{Eq: RGE}
\end{align}
where $g_s$ is the strong coupling and the electroweak interactions have been neglected. As can be seen in Eq.~\ref{Eq: yp}, the value of $y'$ for $\mu = m_{\mathrm{top}}$ strongly depends on $\kappa_t$ and $\kappa_V$. The left-hand side of Fig.~\ref{Fig: RGE} shows the values of $\kappa_t$ and $\kappa_V$ for which the Yukawa couplings become non perturbative only above $10\TeV$ as well as the intersection of the ATLAS bounds shown in Fig.~\ref{Fig: boundsATLAS}.


Since the ATLAS bounds tend to favor relatively large values of $\kappa_V$, they are in tension with the perturbativity bounds. The intersection of all the allowed ($\kappa_t, \kappa_V$) regions leaves open only a small negative $\kappa_t$ region of the parameter space. Since the allowed regions have been obtained by intersecting several two sigma bounds, a more rigorous statistical analysis might exclude the remaining negative top Yukawa region at two sigma. In what follows, however, we are going to assume that the two regions shown in Fig.~\ref{Fig: RGE} are still allowed by the ATLAS results.

The allowed region of the $(\kappa_t, \kappa_V)$ space corresponding to negative values of the top Yukawa coupling is extremely small and centered around
\begin{align}
    \kappa_t &= -0.9\quad\kappa_V = 0.5
\end{align}
which corresponds to a factor of four suppression of the Higgs production cross section for vector boson fusion and associated production. The detection rate associated to the $W/Z h\rightarrow b\bar b$ searches will therefore be drastically reduced compared to its expected SM value. Although the observed signal strengths for this process in ATLAS  and CMS are consistent with zero \cite{ATLAS:2013sn,Chatrchyan:2013zna}, the bounds from the combined CDF and D0 analysis  at the Tevatron \cite{Aaltonen:2013ipa} are tighter and might lead to further constraints on $\kappa_V$.

The combination of the CDF and D0 results at the Tevatron shows a $3.5$ sigma excess in the $W/Z h\rightarrow b\bar b$ channel \cite{Aaltonen:2013ipa}. Such an excess allows to exclude the null hypothesis at two sigma and therefore, allows to set a two-sigma lower bound on $\kappa_V$. Since the only decay widths that are not proportional to $\kappa_V$ are for $h\rightarrow\gamma\gamma$ and $h\rightarrow gg$ and are subdominant, therefore, the Higgs branching ratios are approximately unaltered. The signal strength associated to our model would then be
\begin{align}
    \mu &= \kappa_V^2 \dfrac{\mathrm{Br}(h\rightarrow b \bar b)}{\mathrm{Br}_{\mathrm{SM}}(h\rightarrow b \bar b)}\sim \kappa_V^2
\end{align}
which leads to
\begin{align}
    \kappa_V \ge \frac{1}{\sqrt{3}}
\end{align}
The exact two-sigma bound in the $(\kappa_t, \kappa_V)$ space is shown in the right-hand side of Fig.~\ref{Fig: RGE}, superimposed on the bounds previously obtained from the perturbativity requirement and from the ATLAS Higgs searches. The Tevatron bound is found to exclude the region of parameter space still allowed by these previous limits.

A minimal model with two Higgses as the one shown in Eq.~\ref{Eq: lagrangian}, with only one of the Higgses getting a vev, cannot explain the negative values of the top Yukawa couplings allowed by ATLAS. One of the main issues of such a model is that, in order for the top Yukawa couplings to remain perturbative up to a high scale, the mixing angle between the two Higgses needs to be large. Such a large mixing results in a significant suppression of the Higgs trilinear coupling to vector bosons, which is incompatible with the current LHC \cite{ATLAS:2013sla} and Tevatron \cite{Aaltonen:2013ipa} Higgs results. Since the different allowed regions are close to each other, we further explore alternate more elaborate models to check how robust our conclusion is. Ultimately, obtaining a model which survives the ATLAS and Tevatron bounds will require adding novel Higgses, as will be shown in Sec.~\ref{Sec: GM}.

\subsection{Possible extensions of the minimal model}
\label{Sec: extensions}

The minimal model described by Eq.~\ref{Eq: lagrangian} excludes any large deviation from the SM top Yukawa couplings. In what follows, we argue that introducing new Higgs couplings as well as choosing more general parameterizations does not open new negative top Yukawa regions.

\begin{figure}
    \includegraphics[width=0.32\linewidth]{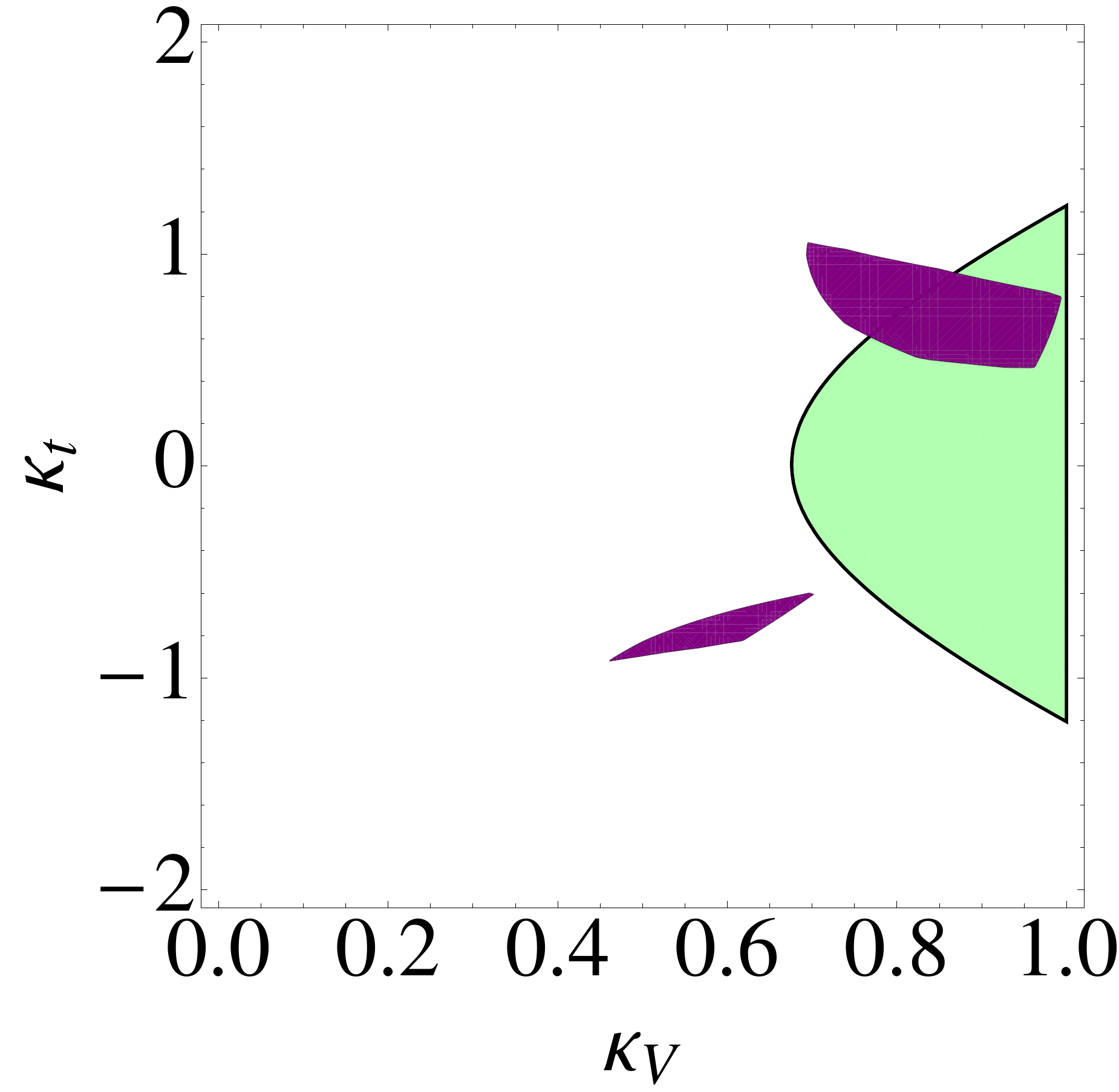}
    \includegraphics[width=0.32\linewidth]{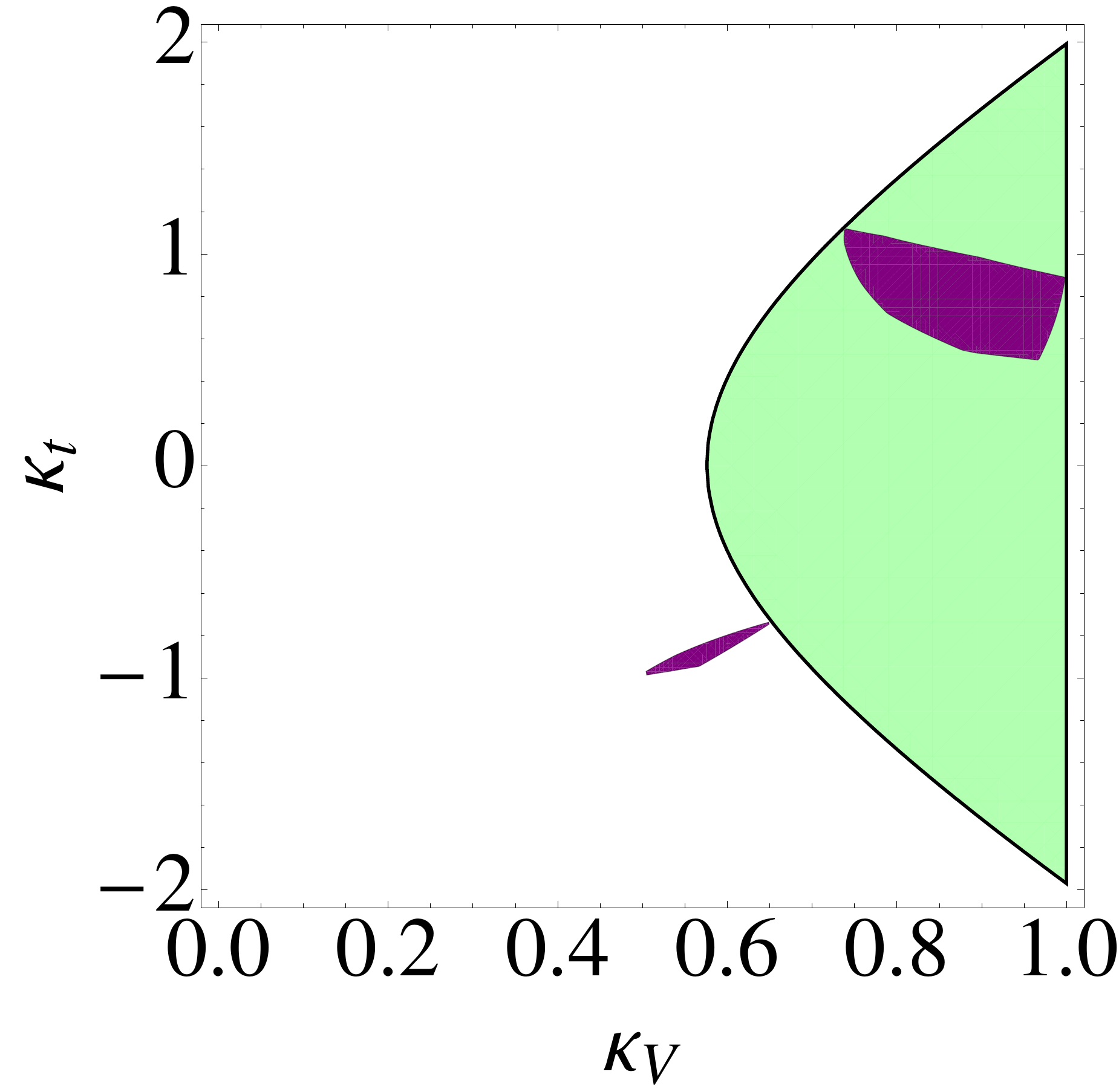}
    \includegraphics[width=0.32\linewidth]{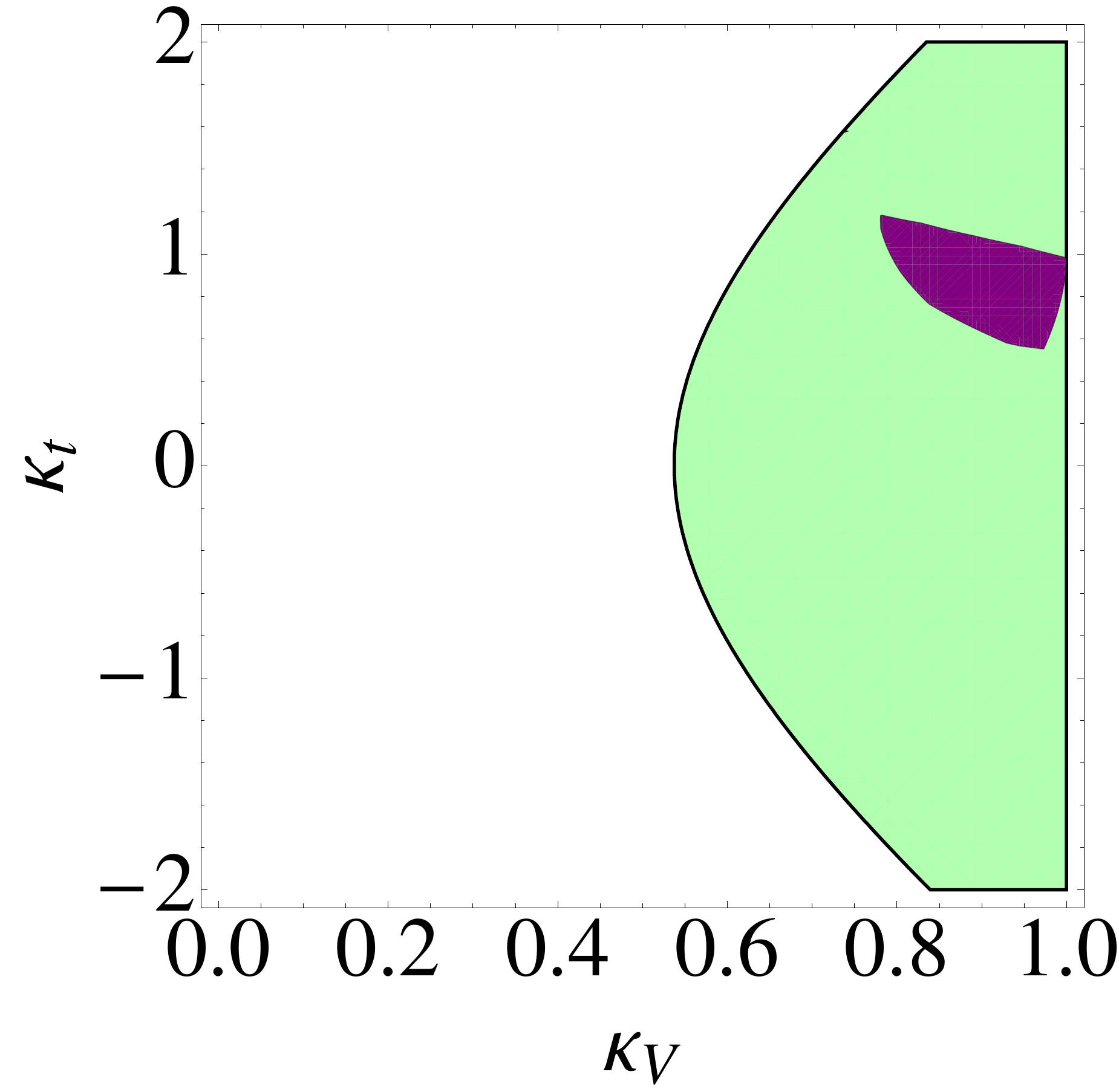}
    \caption{ATLAS and perturbativity bounds (purple) and Tevatron bounds (green) on $\kappa_t$ and $\kappa_V$ with $\kappa_b=0.4$ (left), $\kappa_b=0.5$ (center) and $\kappa_b=0.6$ (right). $\kappa_b\sim 0.5$ is where the different bounds are the closest to each other. \label{Fig: withkb}.}
\end{figure}

The minimal model in Eq.~\ref{Eq: lagrangian} had two free parameters that modified the Higgs couplings at tree level: $\kappa_t$ and $\kappa_V$. Introducing a coupling of $h_2$ to down-type quarks would allow to freely modify the coupling of the Higgs to bottom quarks and have a third handle, $\kappa_b$, defined the same way as $\kappa_t$. If $\kappa_V = \cos\theta$ as before, the perturbativity bounds shown in Fig.~\ref{Fig: RGE} will remain the same. As shown in Fig.~\ref{Fig: withkb}, increasing $\kappa_b$ would lead to weaker Tevatron bounds but also to a uniform suppression of the Higgs detection rates in the different ATLAS Higgs detection channels, due to the lower corresponding branching ratios. Decreasing $\kappa_b$ allows larger detection rates in all the ATLAS channels \cite{ATLAS:2013oma,ATLAS:2013wla,ATLAS:2013nma} but leads to even tighter Tevatron bounds  \cite{Aaltonen:2013ipa}. Both configurations therefore exclude the negative $\kappa_t$ region at two sigma.

It is worth noting that models where both Higgses get a vev can be treated as a subset of the free $\kappa_b$ case described above. Indeed, we can now redefine $\kappa_V$ as
\begin{align}
    \kappa_V = \cos(\theta - \beta)~,
\end{align}
where $\beta$ is the mixing angle between the two vevs such that $\beta = 0$ corresponds to the minimal model studied in Sec.~\ref{Sec: minimal}. The only couplings which will be modified in the $(\kappa_t, \kappa_V)$ space will be the couplings of the Higgs to fermions other than the top, for which we can define
\begin{align}
    \kappa_f = \cos\theta\cos\beta
\end{align}
In practice, since the ATLAS and CMS bounds on the Higgs couplings to fermions other than the top and bottom quark are extremely loose \cite{ATLAS:2013sn,Chatrchyan:2013zna,ATLAS:2013di,ATLAS:2013cs,CMS:2013zz,Chatrchyan:2012vp}, introducing a non zero $\beta$ angle is equivalent to introducing a $\kappa_b$ such that
\begin{align}
    \kappa_b = 
    \begin{cases}
        \kappa_V \cos\beta + \sqrt{1-\kappa_V}\sin\beta\text{\quad for }\theta < \beta\\
        \kappa_V \cos\beta - \sqrt{1-\kappa_V}\sin\beta\text{\quad for }\theta > \beta
    \end{cases}
\end{align}
Models where both Higgses get a vev will then not allow for negative values of the top Yukawa coupling.

As shown in the previous section, the main source of tension between the different Higgs results for a negative top Yukawa is the high suppression of the Higgs coupling to vector bosons. As shown above, increasing the corresponding branching ratios by suppressing the Higgs coupling to bottom quarks leads to a conflict with the Tevatron bounds. These results, however, were obtained for models with only two Higgs doublets. The following sections study how the collider and perturbativity bounds evolve when additional scalars are introduced.

\subsection{Arbitrary number of Higgs doublets}
\label{Sec: Nhiggs}
The conclusions of the previous sections were valid for two Higgs doublets. Here, we argue that they can be extended to the case of an arbitrary number of $SU(2)$ doublets mixing with each other. In particular, we show that adding Higgs doublets makes the perturbativity bounds tighter.

The RGEs for the Yukawa couplings \cite{Machacek:1983fi} remain similar to the ones in Eq.~\ref{Eq: RGE}
\begin{align}
    16\pi^2\frac{d\log y_i}{dt} &= \frac{9}{2}\sum_{k=1\ldots N} y_k^2 - 8 g_s^2
    \label{Eq: RGEN}
\end{align}
Since the right hand side of Eq.~\ref{Eq: RGEN} is the same for all couplings, the ratios of the Yukawa couplings do not depend on the energy scale. Writing $y_i(t) = y_1(t) K_i$ gives
\begin{align}
    16\pi^2 \frac{d y_1}{dt} &= \frac{9}{2}y_1^3 \sum_{k=1\ldots N} K_k^2 -8 y_1 g_s^2
\end{align}
Close to the Landau pole, the $y_1^3$ term will dominate and the RGE can be solved analytically. The position of the Landau pole, $\mu_L$, will be given by
\begin{align}
    \sum_{k=1\ldots N} y_k^2\log\frac{\mu_L}{\mu_0} &= \frac{16\pi^2}{9}
\end{align}
Requiring no Landau poles below $10\TeV$ leads to the following constraint on the sum of the squares of the Yukawa couplings
\begin{align}
    \sum_{k=1\ldots N} y_k^2 &\le 4.33
\end{align}
which can be rewritten as
\begin{align}
    |\vec y|^2 &\le 4.33
    \label{Eq: bound}
\end{align}
by defining $\vec y  = (y_1, \ldots, y_{N})$. Similarly, defining the vev as $\vec v = (v_1, \ldots, v_N)$ and the Higgs mass eigenstates as $\vec H = (H_1, \ldots, H_N)$ in the orthonormal basis defined by the Higgs interaction eigenstates, the constraints on the initial values of the top Yukawas are
\begin{align}
    \vec y . \hat v &= y_{SM},\quad\vec y . \hat H = \kappa_t y_{SM}
    \label{Eq: init}
\end{align}
Redefining $\hat H$ and $\vec y$ as
\begin{align}
    \hat H &= \cos\theta \hat v + \sin\theta \hat H_\perp,\quad\vec y = y_{SM}\hat v + \vec y_\perp
\end{align}
Eq.~\ref{Eq: init} becomes
\begin{align}
    \vec y_\perp . \hat H_\perp = y_{SM}\frac{\kappa_t  - \cos\theta}{\sin\theta}
\end{align}
Thus,
\begin{align}
|\vec y_\perp| \ge y_{SM}\frac{|\kappa_t  - \cos\theta|}{\sin\theta}
\end{align}
So we finally get
\begin{align}
    |\vec y|^2 \ge \frac{1+\kappa_t^2 - 2 \kappa_t\cos\theta}{\sin^2\theta} 
\end{align}
Writing the vector boson couplings after symmetry breaking gives
\begin{align}
    \kappa_V = \cos\theta
\end{align}
and finally, we get
\begin{align}
    |\vec y|^2 \ge \frac{1+\kappa_t^2 - 2 \kappa_t\kappa_V}{1-\kappa_V^2} 
    \label{Eq: Ylower}
\end{align}
Combining this result with Eq.~\ref{Eq: bound}, we see that a \emph{necessary} condition to avoid Landau poles below $10\TeV$ is 
\begin{align}
    \frac{1+\kappa_t^2 - 2 \kappa_t\kappa_V}{1-\kappa_V^2} &\le 4.33
    \label{Eq: necCond}
\end{align}
The bound in Eq.~\ref{Eq: Ylower} is saturated if there are two Higgses. The bounds on $\kappa_t$ and $\kappa_V$ associated to Eq.~\ref{Eq: necCond} are then the ones shown in Fig.~\ref{Fig: RGE}. For more than two Higgses, the bounds in Eq.~\ref{Eq: Ylower} are not saturated and the associated bounds on $\kappa_t$, $\kappa_V$ are more restrictive than the ones shown in Fig.~\ref{Fig: RGE}. So if negative values of $\kappa_t$ are not allowed for two Higgs bosons, they are also not allowed for any larger number of Higgs bosons. Note that this result also applies when all Higgs scalars get arbitrary vevs.

SM extensions with an arbitrary number of Higgs doublets and a negative top Yukawa coupling are then either excluded by ATLAS and the Tevatron, or have low scale Landau poles. The tension between the different collider bounds is caused by the large mixing angle suppression of the Higgs coupling to vector bosons required by the perturbativity bounds. Viable extensions of the minimal model studied in Sec.~\ref{Sec: minimal} should then have $\kappa_V$ close to one. 
\subsection{New representations of $SU(2)$}
\label{Sec: GM}
The $125\GeV$ Higgs boson observed in the LHC couples to vector bosons and fermions while the second Higgs boson introduced in our model has to couple at least to the top quark to drive the top Yukawa negative. Therefore, both Higgses have to be SU(2) doublets, and their couplings to gauge bosons are then fully determined. The only way to increase $\kappa_V$ would then be to mix the two Higgses with scalars embedded in higher representations of $SU(2)$ \cite{Akeroyd:1900zz,Yagyu:2012qp,Yagyu:2013kva,Hisano:2013sn,Georgi:1985nv}. In what follows, we focus on representations of isospin less than two, so triplet scalars.

Introducing a single scalar triplet, $\Phi$, is the most straightforward way of extending our model \cite{Godfrey:2010qb,Akeroyd:1900zz,Yagyu:2012qp,Yagyu:2013kva}. This triplet gets a vev, $v_\Delta$, and includes a neutral scalar particle $\phi^0$, which mixes with the two Higgses. The mixing angles and the vevs are defined such that
\begin{align}
    h^0 &= h^0_1 \cos\alpha \cos\theta + h^0_2\cos\alpha \sin\theta + \phi^0\cos\gamma\\
    v_\Delta &= v\sin\alpha
\end{align}
Defining $\beta$ as the angle between the vevs of the two Higgs doublets, the ratios of the $125\GeV$ Higgs couplings to gauge bosons over their SM values are
\begin{align}
    \kappa_{hWW} &= \cos\alpha\sin\gamma\cos(\beta - \theta) + \sqrt{2(2 - Y^2)}\sin\alpha\cos\gamma\\
    \kappa_{hZZ} &= \frac{\cos\alpha\sin\gamma\cos(\beta - \theta) + 4 Y^2 \sin\alpha\cos\gamma}{\sqrt{\cos^2\alpha + 4Y^2 \cos^2\alpha}}
\end{align}
where $Y$ is the hypercharge of the Higgs triplet. The couplings of the $125\GeV$ Higgs to fermions other than the top are then defined by
\begin{align}
    \kappa_{hff} = \frac{\sin\gamma}{\cos\alpha}\frac{\cos\theta}{\cos\beta}
\end{align}
For $Y \le 1$, a new Higgs triplet could then in principle increase $\kappa_V$ without leading to low scale Landau poles. However, since such triplet models violate custodial symmetry, $v_\Delta$ is strongly constrained to be \cite{Beringer:1900zz}:
\begin{align}
    v_\Delta < 3\GeV
\end{align}
which leads to 
\begin{align}
    \sin\alpha < 0.01
\end{align}
and the resulting increase in $\kappa_V$ will be negligible.

In order to get a significant increase in $\kappa_V$, the additional scalars have to come in complete $SU(2)_C$ multiplets. This can be realized by introducing one $Y=1$ complex triplet and one $Y=0$ triplet, which can be rearranged to form a $(3,\bar 3)$ of $SU(2)_L\times SU(2)_R$:
\begin{align}
    \Delta = \begin{pmatrix}
        \chi^{0*} & \xi^+ & \chi^{++}\\
        \chi^- & \xi^0 & \xi^+\\
        \chi^{--} & \xi^- & \chi^0
    \end{pmatrix}
\end{align}
This model can be viewed as a simple extension of the Georgi-Machacek model \cite{Georgi:1985nv}, with one additional Higgs doublet. In order for custodial symmetry to be enforced, the triplet vevs have to be diagonal
\begin{align}
    v_\chi = v_\xi
\end{align}
This time, the mixing angle $\alpha$ is defined such that
\begin{align}
    v_\chi = \frac{v}{\sqrt{8}}\sin\alpha
\end{align}
and the $125\GeV$ Higgs couplings become
\begin{align}
    \kappa_{hWW} &= \kappa_{hZZ} = \kappa_V =  \cos\alpha\sin\gamma\cos(\beta - \theta) + \sqrt{\frac{8}{3}}\sin\alpha\cos\gamma\\
    \kappa_{hff} &= \frac{\sin\gamma}{\cos\alpha}\frac{\cos\theta}{\cos\beta}
\end{align}
Since this model does not violate custodial symmetry at tree level, the triplet vev $v_\chi$ can now go up to about $100\GeV$. Such a large vev allows to alleviate the tension between the ATLAS and Tevatron bounds while avoiding low scale Landau poles. 
 
\begin{figure}
    \centering
    \includegraphics[width=0.5\linewidth]{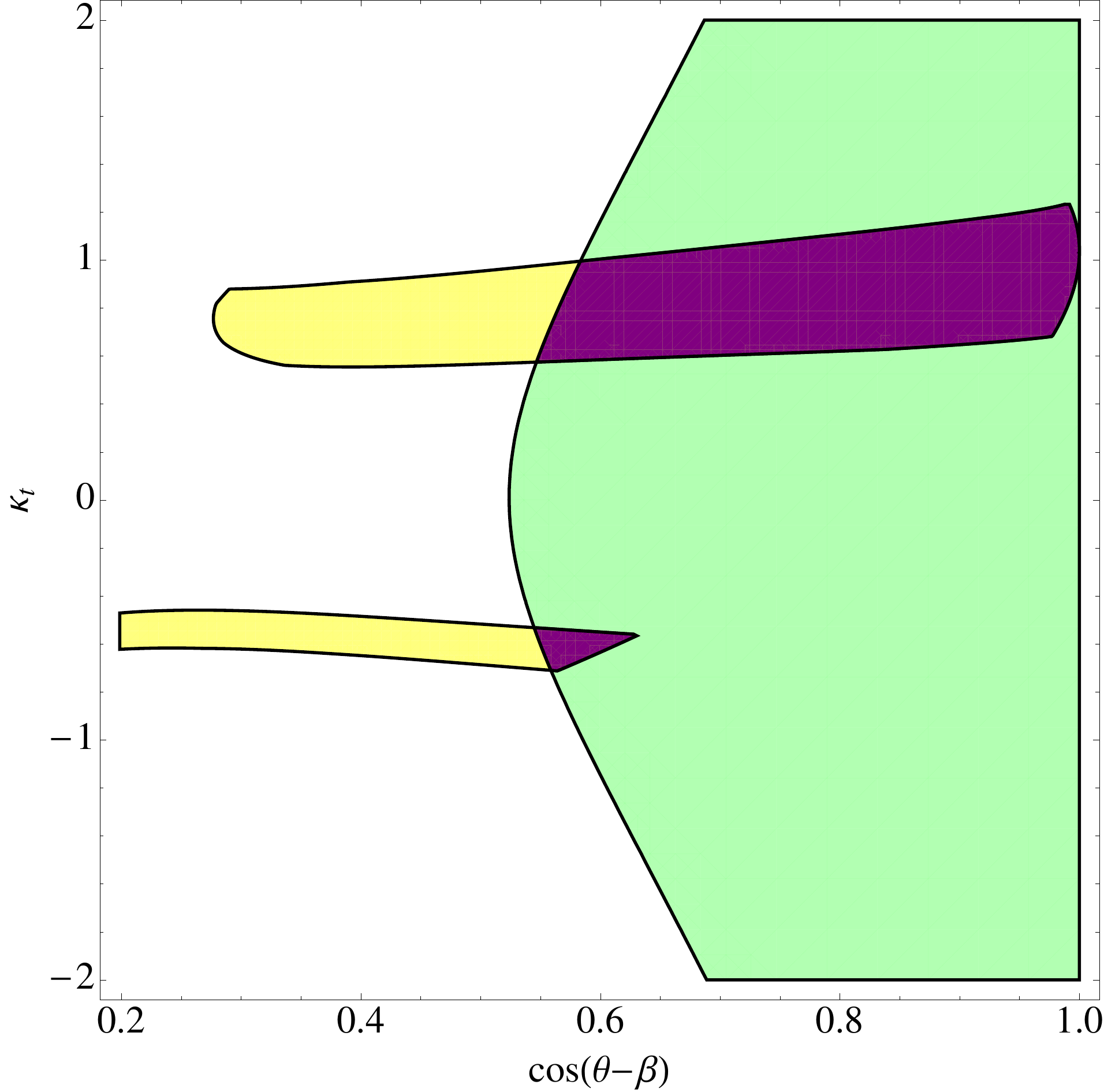}
    \caption{Perturbativity bounds and ATLAS $h\rightarrow WW\rightarrow ll\nu\nu$ \cite{ATLAS:2013wla}, $h\rightarrow \gamma \gamma$ \cite{ATLAS:2013oma} and $h\rightarrow ZZ \rightarrow 4l$   \cite{ATLAS:2013nma} bounds (yellow) in the $(\cos(\theta - \beta), \kappa_t)$ space. The Tevatron \cite{Aaltonen:2013ipa} bound is shown in green. The intersection of all the different bounds is shown in purple. The model is a Georgi-Machacek \cite{Georgi:1985nv} model with two Higgs doublets instead of one and $\alpha = 0.52$, $\beta = 0.2$ and $\gamma = 1.12$.\label{Fig: GM}}
\end{figure}

Since this model respects the custodial symmetry $SU(2)_C$ \cite{Chiang:2012cn}, the nine Higgs fields in $\Delta$, can be rearranged in full $SU(2)_C$ multiplets -- a five-plet $H_5$, a triplet $\tilde{H}_3$ and a singlet $\tilde{H}_1$ -- using
\begin{align}
    \mathbf{9} = \mathbf{5} \oplus \mathbf{3} \oplus \mathbf{1}
\end{align}
Besides the $125\GeV$ Higgs $h^0$, this model then implies the existence of $12$ new scalars:
\begin{itemize}
\item 3 Goldstone bosons which are eaten by the $W$ and $Z$ bosons
\item 3 charged Higgses $H_1^\pm$, $H_2^\pm$ and $H_3^\pm$
\item one doubly charged Higgs $H_3^{++}$
\item 3 pseudoscalar Higgses $A^0_1$, $A^0_2$ and $A^0_3$
\item 2 CP-even Higgses $H^0_1$ and $H^0_2$
\end{itemize}
Fig.~\ref{Fig: GM} shows the ATLAS, Tevatron and perturbativity bounds for $\alpha = 0.52$, $\beta = 0.20$ and $\gamma = 1.12$ in function of $\kappa_t$ and $\cos(\theta - \beta)$. Although the tension between the different ATLAS  \cite{ATLAS:2013oma,ATLAS:2013wla,ATLAS:2013nma} and Tevatron \cite{Aaltonen:2013ipa} searches is alleviated in this new model, the allowed region of parameter space remains limited.

The Georgi-Machacek model \cite{Georgi:1985nv} with one additional Higgs doublet allows us to reconcile the different collider and perturbativity bounds on the $125\GeV$ Higgs boson. It predicts, however, a large number of new particles which are within the reach of the current LHC searches.
\subsubsection{The need for alternate Higgs decay modes}
The Georgi-Machacek model studied here predicts the existence of three neutral CP-even Higgs bosons: a $125\GeV$ Higgs analogous to the one observed in ATLAS and CMS, $h^0$, and two new scalars $H_1^0$ and $H_2^0$, whose masses will be constrained by the current ATLAS and CMS Higgs results. For Higgses heavier than about $140\GeV$, the most sensitive search is the CMS $H\rightarrow ZZ\rightarrow 4l$ search \cite{CMS:2012xaa}, which constrains the detection rate of heavy Higgses to be no larger than about $30$\% of the SM one. Lighter Higgs bosons are constrained by the ATLAS $H\rightarrow \gamma\gamma$ search \cite{ATLAS:2013oma}, and, in most of the mass range, their detection rate in this channel should be no larger than about half the SM one. Here again, we choose to leave apart the region where the two Higgses are nearly degenerate.

Although both ATLAS and CMS require $H_1^0$ and $H_2^0$ to have suppressed couplings, 
generating a large negative Yukawa coupling requires one of the Higgs interaction eigenstates to have a large coupling $y'$ to the top, as shown in Fig.~\ref{Fig: ypkv} and Eq.~\ref{Eq: yp}. In the minimal model shown in Sec.~\ref{Sec: minimal}, this large $y'$ leads to an enhanced coupling of the second Higgs mass eigenstate to the top quark
\begin{align}
    \frac{y_{Ht\bar t}}{y_{SM}} &= \frac{\kappa_t \kappa_V - 1}{\sqrt{1-\kappa_V^2}} > 1\text{ for }\kappa_t < 0.
\end{align}
This enhanced coupling to the top quark leads to a larger production rate of the second Higgs boson through gluon fusion. Here, for $\kappa_t = -1$ and $\kappa_V = 0.5$, we would have 
\begin{align}
    \frac{\sigma_{GGF}}{\sigma_{GGF}^{SM}} = \left(\frac{\kappa_t \kappa_V - 1}{\sqrt{1-\kappa_V^2}}\right)^2 = 3.
\end{align}
In order for $H_1^0$ and $H_2^0$ to be within the current ATLAS and CMS bounds, the branching ratios in the channels of interest should then be reduced by a factor of about $10$. In order to our models to be viable, the new Higgs bosons must then couple to particles other than SM fermions and vector bosons.

\begin{figure}
    \centering
    \includegraphics[width=0.5\linewidth]{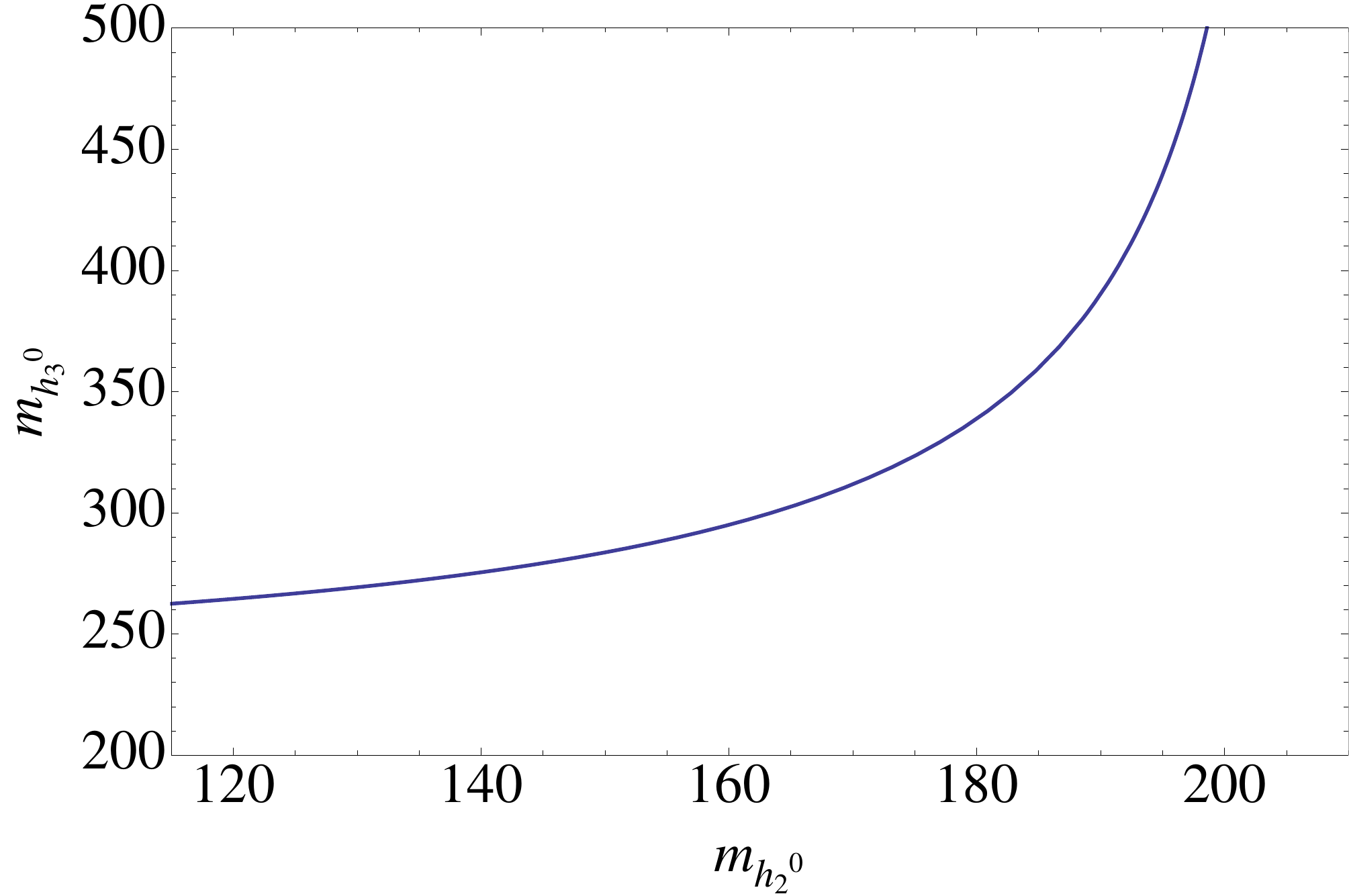}
    \caption{\label{Fig: mphiH}Mass of $\phi^0$ in function of $m_H$ for $\alpha = 0.52$, $\gamma = 1.12$, $\beta = 0.2$ and $\kappa_V = 0.57$.}
\end{figure}

The modified Georgi-Machacek model described here predicts the existence of two new Higgs mass eigenstates, $H_1^0$ and $H_2^0$. In order to drive the top Yukawa coupling negative without significantly suppressing the $125\GeV$ Higgs coupling to vector bosons, the mixing between the three neutral Higgs interaction eigenstates needs to be large. This large mixing constrains the lightest new mass eigenstate, $H_1^0$, to be lighter than $200\GeV$, as shown in Fig.~\ref{Fig: mphiH}. 

Having large mixing angles between the neutral CP-even Higgses also means that all three mass eigenstates $h^0$, $H_1^0$ and $H_2^0$ will have sizeable couplings to the top quark. The production cross sections of $H_1^0$ and $H_2^0$ through gluon fusion will then be comparable to the SM ones. Since these new Higgses have not been observed, they must have non SM decay modes. Although $H^0_2$, if heavy enough, can decay to two Higgs bosons, this decay mode is forbidden for $H_1^0$, which has to be lighter than $200\GeV$. $H^0_1$ then needs to have alternate decay modes. In what follows, we assume that $H^0_1$ and $H^0_2$ can decay to invisible particles. Scenarios involving similar interactions have been studied in models with Higgs-portal dark matter \cite{Silveira:1985rk,McDonald:1993ex,Burgess:2000yq,Davoudiasl:2004be,Patt:2006fw,Andreas:2010dz,Raidal:2011xk,He:2011de,Drozd:2011aa,Mambrini:2011ik}. 

Coupling the Higgs interaction eigenstates to invisible particles can induce new decay modes for the $125\GeV$ Higgs boson. These new invisible decay modes would lead to suppressed detection rates for the $125\GeV$ Higgs boson in all the different detection channels. Our previous two Higgs doublet model, however, suffered from a suppressed coupling of the $125\GeV$ Higgs to vector bosons, which lead to a strong suppression of the detection rate in the $H\rightarrow ZZ\rightarrow 4l$  \cite{ATLAS:2013nma} and the $H\rightarrow WW\rightarrow 2l2\nu$  \cite{ATLAS:2013wla} channels. Since, as shown in Fig.~\ref{Fig: GM}, introducing new Higgs triplets barely compensates for this suppression, opening new invisible decay channels for the $125\GeV$ Higgs would lead us to the same issues as the ones considered in Sec.~\ref{Sec: minimal}. The masses of a new invisible particle $S$ coupling to the neutral Higgs bosons of our model would then be constrained by:
\begin{align}
    \frac{m_h}{2} \sim 63\GeV \lsim m_{S} \lsim \frac{m_{\phi^0}}{2}\lsim 100\GeV.
\end{align}
In what follows, we assume that $S$ is a SM singlet and couples to the Higgs interaction eigenstates through
\begin{align}
    \mathcal{L}&\supset \lambda_1 h_1^\dagger h_1 S^\dagger S + \lambda_2 h_2^\dagger h_2 S^\dagger S + \lambda_3 \Delta^\dagger \Delta S^\dagger S
\end{align}
\subsubsection{A possible model}
\label{Sec: benchmark}
\begin{figure}
\centering
\includegraphics[width=0.7\linewidth]{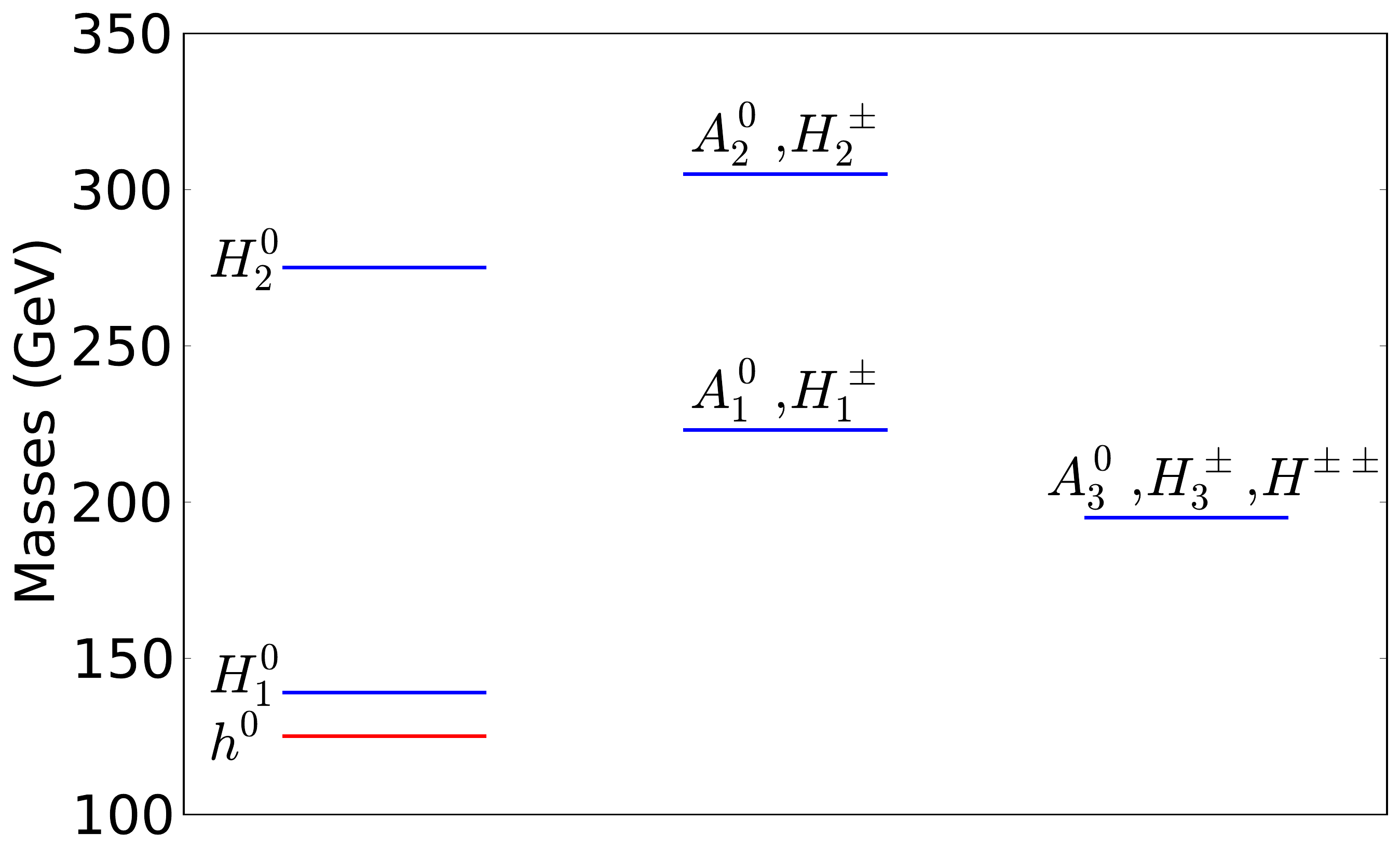}
\caption{\label{Fig: spectrum} Higgs masses for the benchmark model described in Sec.~\ref{Sec: benchmark}. The observed $125\GeV$ Higgs is shown in red.}
\end{figure}

This section studies one possible benchmark model satisfying all the current constraints from collider searches and perturbativity requirements described in Sec.~\ref{Sec: minimal}. The mixing angles we choose here are the same as the ones in Fig.~\ref{Fig: GM}:
\begin{align}
    \alpha = 0.52\quad\beta = 0.2\quad\theta = 1.16\quad \gamma = 1.12
\end{align}
For this choice of angles and 
\begin{align}
\kappa_t = -0.55 \quad\quad \kappa_V = 0.57
\end{align}
one possible spectrum is  
\begin{align}
    m_{H_1^\pm} &= m_{A_1^0} = 223\GeV\\
     m_{H_2^\pm} &= m_{A_2^0} = 305\GeV\\
    m_{H^{++}}&= m_{A_3^0} = m_{H_3^\pm} = 195\GeV\\
m_{H_1^0} &= 139\GeV \quad m_{H_2^0} = 275\GeV
\end{align}
as shown in Fig.~\ref{Fig: spectrum}.
This model satisfies the perturbativity and vacuum stability constraints up to $\Lambda \sim 7\TeV$. Details about the corresponding quartic couplings can be found in Appendix \ref{App: GM}. 

The branching ratios for the $125\GeV$ Higgs become:
\begin{align}
    \Br(h\rightarrow bb) = 0.34\quad \Br(h\rightarrow WW) = 0.46\\
    \Br(h\rightarrow ZZ) = 0.056\quad\Br(h\rightarrow\gamma\gamma) = 0.012
\end{align}
The production cross sections in the gluon and vector boson fusion channels are
\begin{align}
    \sigma(gg\rightarrow h) = 5.9\pb\quad\sigma(VV\rightarrow h) = 1.0\pb
\end{align}
and the signal strength in the $h\rightarrow ZZ\rightarrow 4l$ and the $h\rightarrow\gamma\gamma$ channels are
\begin{align}
    \mu(h\rightarrow ZZ\rightarrow 4l) = 0.70\quad\mu(h\rightarrow\gamma\gamma) = 1.67
\end{align}
The corresponding signal strengths for $H_1^0$ and $H_2^0$ when no alternate decay modes are introduced are
\begin{align}
    \mu(H_1\rightarrow ZZ\rightarrow 4l) &= 2.1\quad\mu(H_1\rightarrow\gamma\gamma) = 1.46\\
    \mu(H_2\rightarrow ZZ\rightarrow 4l) &= 0.47
\end{align}
while the corresponding ATLAS  \cite{ATLAS:2013oma} and CMS \cite{CMS:2012xaa} upper bounds are
\begin{align}
    \mu^{\mathrm{CMS}}_{\mathrm{ZZ}}(m_{H_1^0})& ~< 0.13\quad \mu^{\mathrm{ATLAS}}_{\mathrm{\gamma\gamma}}(m_{H_1^0})~< 0.74\\
    \mu^{\mathrm{CMS}}_{\mathrm{ZZ}}(m_{H_2^0})& ~< 0.30
\end{align}
Introducing a SM scalar singlet $S$, which couples to the $h_1$ interaction eigenstate via the operator
\begin{align}
    \mathcal{L} &= \lambda_{1} h_1^\dagger h_1 S^\dagger S
\end{align}
allows to reduce the different signal strength down to below the ATLAS and CMS bounds. For 
\begin{align}
    \lambda_{1} = 0.2\quad m_{S} = 66\GeV,
\end{align}
the signal strengths for $H_1^0$ and $H_2^0$ become
\begin{align}
    \mu(H_1\rightarrow ZZ\rightarrow 4l) &= 0.024\quad\mu(H_1\rightarrow\gamma\gamma) = 0.017\\
    \mu(H_2\rightarrow ZZ\rightarrow 4l) &= 0.25
\end{align}
The charged Higgs are dominantly produced in association with a top quark through $gb\rightarrow tH^-$, with cross sections of the order of the picobarn, and dominantly decay to $tb$ and $Wh$ \cite{Ferrari:2013pma,Flechl:2013kaa}. Since the decay to $tb$ suffers from high background rates, current searches focus on rare decays to $\tau\nu$ \cite{Saavedra:2013ina} and $cs$ \cite{Aad:2013hla}, giving weak exclusion bounds. Since these rare decays are further suppressed in our model by the fact that only one Higgs interaction eigenstate couples to leptons and light quarks, current charged Higgs searches are far from excluding our benchmark model. Associated production of a charged Higgs with a top quark can lead to final states of the form $t\bar t b$ and could therefore be probed using $t\bar t$ cross section measurements. However, the current uncertainties set by ATLAS and CMS on the top pair production cross section at $8\TeV$ are of the order of $10\pb$ and deviations due to the associated production of charged Higgses would not be visible \cite{Aad:2013hpa,CMS:2013jea}. Pair production of charged Higgses occurs only through electroweak processes, with cross sections of the order of $10\fb$. Although these processes lead to more characteristic final states, SM backgrounds involving top quarks and gauge bosons would be largely dominating.

The doubly-charged Higgs can decay only to a pair of same-sign $W$ bosons, which can produce signals with same-sign lepton pairs. Searches for such signatures would have particularly low background rates, and could lead to strong constraints on the doubly-charged Higgs mass, in spite of its low production rate. Existing LHC  \cite{ATLAS:2012hi,CMS:2012ulp} and Tevatron  \cite{Acosta:2004uj,Abazov:2011xx} searches for these particles, however, consider only direct decays to same-sign lepton pairs. Recasting these searches to look for doubly-charged Higgses decaying to vector bosons has been done in  \cite{Kanemura:2013vxa} and gives only weak bounds. Since the doubly-charged Higgs does not couple to leptons, it would be dominantly produced electroweakly, in association with a charged Higgs, with a cross section of the order of $10\pb$. The corresponding final states are similar to the ones probed by the CMS and ATLAS SUSY searches for a same-sign lepton pair produced with jets \cite{CMS:2013jea,Aad:2013hppa}. The number of observed events, however, would be too small to lead to significant excesses in the signal regions of interest.  Dedicated searches for doubly-charged Higgses decaying to $W$ pairs have been designed in \cite{Godfrey:2010qb,Azuelos:2004dm,Chiang:2012dk} and seem to indicate that, for masses comparable to ours, a discovery could be made with the current $8\TeV$ dataset. 

\section{Discussion}
\label{Sec: discussion}

This article investigates the different low energy scenarios that could lead to the negative values of the top Yukawa coupling still allowed by the ATLAS Higgs searches. We focus on three possible scenarios involving either new vector-like top partners or new Higgs-like scalars. In order for the new physics contribution to the top Yukawa to be large enough, the new particles have to be light and have large couplings to the top quark. Scenarios with new vector-like top partners and a negative top Yukawa are therefore either excluded by the current LHC searches or  leading to low scale Landau poles. Viable models with additional Higgs-like scalars have to include additional scalar doublets and triplets which heavily mix with each other. One possible scenario which is still allowed by the current ATLAS and Tevatron Higgs searches is a Georgi-Machacek model with one additional Higgs doublet. This model would predict a large number of new charged and neutral Higgses with sizable couplings to the top quark. Designing new LHC searches or recasting existing ones would allow to strongly constrain the corresponding parameter space, even with the current $8\TeV$ dataset.

\section*{Acknowledgements}{\label{sec:ack}}
SE and JW are supported by the US DOE under contract number DE-AC02-76-SF00515. SE is supported by a Stanford Graduate Fellowship. Fermilab is operated by Fermi Research Alliance, LLC under Contract No. DE-AC02-07CH11359 with the United States Department of Energy.
\appendix
\section{RGEs for the 2HDM}
\label{App: 2hdm}
The Higgs potential for a general two Higgs doublet model (2HDM) is
where
\begin{align}
h_1 = \begin{pmatrix}
G^\pm \cos\beta + H^\pm\sin\beta\\
\dfrac{v\cos\beta + h_1^0 + i (G^0\cos\beta + A^0\sin\beta) }{\sqrt{2}}
\end{pmatrix}\quad
h_2 = \begin{pmatrix}
-G^\pm \sin\beta + H^\pm\cos\beta\\
\dfrac{v\sin\beta + h_2^0 + i (-G^0\sin\beta + A^0\cos\beta) }{\sqrt{2}}
\end{pmatrix}
\end{align}
Both Higgs doublets couple to the top quark
\begin{align}
    \LL_{\mathrm{Yukawa}} &= y_t h Q u^c + y_t' H Q u^c
\end{align}
The most general Higgs potential in the 2HDM can be expressed as
\begin{align}
V &= m_1^2 h_1^\dagger h_1 + m_2^2 h_2^\dagger h_2 + m_{12}^2 (h_1^\dagger h_2 + \mathrm{c.c.})\\
\nonumber
& + \frac{\lambda_1}{2}\left(h_1^\dagger h_1\right)^2 + \frac{\lambda_2}{2} \left(h_2^\dagger h_2\right)^2 + \lambda_3 (h_1^\dagger h_1)(h_2^\dagger h_2) + \lambda_4 (h_1^\dagger h2)(h_2^\dagger h_1)\\
\nonumber
& + \frac{\lambda_5}{2}\left[\left(h_1^\dagger h_2\right)^2 +  \mathrm{c.c.}\right] + \lambda_6 \left[(h_1^\dagger h_1)(h_1^\dagger h_2) + \mathrm{c.c.} \right] +  \lambda_7 \left[(h_2^\dagger h_2)(h_2^\dagger h_1) + \mathrm{c.c.} \right] 
\end{align}
The beta functions for the Yukawa and quartic couplings are then (neglecting the $SU(2)\times U(1)$ couplings)
\begin{align*}
    16\pi^2\beta_{\lambda_1} &= 12 \lambda_1^2 + 4\lambda_3^2 + 4\lambda_3\lambda_4 + 2\lambda_4^2 + 2\lambda_5^2 + 24\lambda_6^2 + 12\lambda_1 y_t^2 + 12\lambda_6 y_ty'_t-12 y_t^4\\
    16\pi^2\beta_{\lambda_2} &= 12 \lambda_2^2 + 4\lambda_3^2 + 4\lambda_3\lambda_4 + 2\lambda_4^2 + 2\lambda_5^2 + 24\lambda_7^2 + 12\lambda_2 y'{}_t^2+ 12\lambda_7 y_ty'_t-12 y'{}_t^4\\
    16\pi^2\beta_{\lambda_3} &= (\lambda_1+\lambda_2)(6\lambda_3 + 2\lambda_4) + 4\lambda_3^2 + 2\lambda_4^2 + 2\lambda_5^2 + 4\lambda_6^2 + 16\lambda_6\lambda_7 + 4\lambda_7^2\\
    &+6\lambda_3(y_t^2 + y'{}_t^2)+6(\lambda_6+\lambda_7)y_ty'_t-12 y_t^2y'{}_t^2\\
    16\pi^2\beta_{\lambda_4} &= 2(\lambda_1+\lambda_2)\lambda_4 + 8\lambda_3\lambda_4 + 4\lambda_4^2 + 8\lambda_5^2 + 10\lambda_6^2 + 4\lambda_6\lambda_7 + 10\lambda_7^2\\
    &+6\lambda_4(y_t^2 + y'{}_t^2)+6(\lambda_6+\lambda_7)y_ty'_t-12 y_t^2y'{}_t^2\\
    16\pi^2\beta_{\lambda_5} &= 2(\lambda_1+\lambda_2)\lambda_5 + 8\lambda_3\lambda_5 + 12\lambda_4\lambda_5 + 10\lambda_6^2 + 4\lambda_6\lambda_7 + 10\lambda_7^2\\
    &+6\lambda_5(y_t^2 + y'{}_t^2)+6(\lambda_6+\lambda_7)y_ty'_t-12 y_t^2y'{}_t^2\\
    16\pi^2\beta_{\lambda_6} &= 12\lambda_1\lambda_6 + 6\lambda_3(\lambda_6+\lambda_7)+8\lambda_4\lambda_6+4\lambda_4\lambda_7+10\lambda_5\lambda_6+2\lambda_5\lambda_7\\
    &+9\lambda_6y_t^2 + 3\lambda_6 y'{}_t^2+3\left(\lambda_1+\lambda_3 + \lambda_4 + \lambda_5\right)y_ty'_t-12 y_t^3y'_t\\
    16\pi^2\beta_{\lambda_7} &= 12\lambda_2\lambda_7 + 6\lambda_3(\lambda_6+\lambda_7)+8\lambda_4\lambda_7+4\lambda_4\lambda_6+10\lambda_5\lambda_7+2\lambda_5\lambda_6\\
    &+9\lambda_7y'{}_t^2 + 3\lambda_7 y_t^2+3\left(\lambda_2+\lambda_3 + \lambda_4 + \lambda_5\right)y_ty'_t-12 y_t^3y'_t\\
    16\pi^2\beta_{y_t} &= y_t\left(\frac{9}{2}y_t^2 + \frac{3}{2}y'{}_t^2 -8 g_3^2\right) + 3y_t y'{}_t^2\\
    16\pi^2\beta_{y'_t} &= y'_t\left(\frac{9}{2}y'{}^2_t + \frac{3}{2}y_t^2 -8 g_3^2\right) + 3y_t^2y'{}_t\\
    16\pi^2\beta_{g_3} &= -7 g_3^3
\end{align*}
The stability conditions are given by \cite{Ferreira:2004yd}
\begin{align}
&\lambda_1 >0, \quad \lambda_2 >0,\quad\lambda_3 > -2\sqrt{\lambda_1 \lambda_2}\\
&\lambda_3 +\frac{\lambda_4+\lambda_5}{2}> -2\sqrt{\lambda_1 \lambda_2}\\
&\lambda_3 +\frac{\lambda_4-\lambda_5}{2}> -2\sqrt{\lambda_1 \lambda_2}\\
&\lambda_1+\lambda_2+\lambda_3+\frac{\lambda_4+\lambda_5}{2}-2|\lambda_6+\lambda_7|>0
\end{align}
\section{The Georgi-Machacek model}
\label{App: GM}
Let us consider the following Higgs potential
\begin{align}
V &= \frac{1}{2} m_{11}\mathcal{H}_{11}^\dagger\mathcal{H}_{11} +  \frac{1}{2} m_{22}\mathcal{H}_{22}^\dagger\mathcal{H}_{22} + \frac{1}{2} m_{12}\mathcal{H}_{12}^\dagger\mathcal{H}_{12} + \frac{1}{8} \lambda_1\left(\mathcal{H}_{11}^\dagger\mathcal{H}_{11}\right)^2\\\
\nonumber
& + \frac{1}{8} \lambda_2\left(\mathcal{H}_{22}^\dagger\mathcal{H}_{22}\right)^2 +  \frac{1}{4} \lambda_3\left(\mathcal{H}_{11}^\dagger\mathcal{H}_{11}\right)\left(\mathcal{H}_{22}^\dagger\mathcal{H}_{22}\right) +  \frac{1}{4} \lambda_4\left(\mathcal{H}_{12}^\dagger\mathcal{H}_{12}\right)\left(\mathcal{H}_{21}^\dagger\mathcal{H}_{21}\right)\\
\nonumber
& +\frac{1}{8} \lambda_5\left(\mathcal{H}_{12}^\dagger\mathcal{H}_{12}\right)^2 
+  \frac{1}{4} \lambda_6\left(\mathcal{H}_{11}^\dagger\mathcal{H}_{11}\right)\left(\mathcal{H}_{12}^\dagger\mathcal{H}_{12}\right) +  \frac{1}{4} \lambda_7\left(\mathcal{H}_{22}^\dagger\mathcal{H}_{22}\right)\left(\mathcal{H}_{12}^\dagger\mathcal{H}_{12}\right)\\
\nonumber
& - \mu \mathrm{Tr}\left(\mathcal{H}_{11}^\dagger \frac{\tau_a}{2}\mathcal{H}_{11}\frac{\tau_b}{2}\right) (P^\dagger\Delta P)^{ab}\\
\nonumber
& + \mathrm{c.c.}
\label{Eq: GMv}
\end{align}
To make the custodial invariance manifest, the Higgs doublets and triplets are represented using the matrices $\mathcal{H}_{ij}$ and $\Delta$ respectively, with 
\begin{align}
\mathcal{H}_{ij} &= \begin{pmatrix}
h_i^{0*} & h_j^+\\
-h_i^- & h_j^0
\end{pmatrix}\quad \quad    \Delta = \begin{pmatrix}
        \chi^{0*} & \xi^+ & \chi^{++}\\
        \chi^- & \xi^0 & \xi^+\\
        \chi^{--} & \xi^- & \chi^0
    \end{pmatrix}
\end{align}
The $\tau_i$ are the Pauli matrices and $P$ is given by
\begin{align}
P &= \begin{pmatrix}
-\frac{i}{\sqrt{2}} & \frac{i}{\sqrt{2}} & 0\\
 \frac{i}{\sqrt{2}} &  -\frac{i}{\sqrt{2}} & 0\\
 0 & 0 & 1
\end{pmatrix}
\end{align}

The fields $h^0_1$, $h^0_2$, $\xi^0$ and $\chi^0$ can take vevs
\begin{align}
h_1^0 &= \frac{v_h\cos\beta + h^0 + i(G^0 \cos\beta + A^0\sin\beta)}{\sqrt{2}}\\
 h_2^0 &= \frac{v_h\sin\beta + H^0 + i(-G^0 \sin\beta + A^0\cos\beta)}{\sqrt{2}}\\
\chi_0 &= v_\chi + \frac{\chi_r + i\chi_i}{\sqrt{2}}, \quad \xi_0 = v_\xi + \xi_r
\end{align} 
We also define
\begin{align}
h_1^\pm &= G^\pm \cos\beta + H_0^\pm \sin\beta\\
h_2^\pm &= H_0^\pm \cos\beta - G^\pm \sin\beta
\end{align}
For our model to remain invariant under custodial symmetry $SU(2)_C$, we need
\begin{align}
v_\xi = v_\chi = v_\Delta
\end{align}
and therefore, we define
\begin{align}
v_\Delta = v\sin\alpha \quad v_h = v\cos\alpha
\end{align}
The nine fields contained in $\Delta$ can be decomposed in one $SU(2)_C$ singlet $H_1$, one triplet $H_3$ and one five-ply $H_5$ defined by
\begin{align}
H_5^{\pm\pm} &= \chi^{\pm\pm} \quad H_5^\pm = \dfrac{1}{\sqrt{2}}(\chi^\pm - \xi^\pm) \quad H^0_5 = \dfrac{1}{\sqrt{3}}(\chi_r-\sqrt{2}\xi_r)\\
H_3^\pm &= \dfrac{1}{\sqrt{2}} (\chi^\pm + \xi^\pm) \quad H^0_3 = \chi_i \quad H^0_1 = \dfrac{1}{\sqrt{3}}(\xi_r + \sqrt{2}\chi_r)
\end{align}
The members of the $SU(2)_C$ five-plet $H_5$ do not mix with any other field and their mass is
\begin{align}
m_{5} &= \frac{\mu v_h^2\cos^2\beta}{4v_\Delta}
\end{align}
The mass matrices for the other fields are given in the $(H_1^0,h_1^0, h_2^0)$ basis for the CP-even scalars, $(H_3^0, G^0,A^0)$ basis for the CP-odd scalars, and $(H_3^\pm,G^\pm,H_0^\pm)$ basis for the charged scalars 
\begin{align}
M_{\mathrm{charged}} = \begin{pmatrix}
\dfrac{\mu v_h^2\cos^2\beta}{4v_\Delta}& -\dfrac{v_h\mu\cos^2\beta}{\sqrt{2}} & \dfrac{v_h\mu\cos\beta\sin\beta}{\sqrt{2}} \\
-\dfrac{v_h\mu\cos^2\beta}{\sqrt{2}} & 2 v_\Delta\mu\cos^2\beta & -v_\Delta\mu\sin(2\beta)\\
 \dfrac{v_h\mu\cos\beta\sin\beta}{\sqrt{2}}& -v_\Delta\mu\sin(2\beta) & m_{H^\pm_0}^2
\end{pmatrix}
\end{align}
\begin{align}
M_{\mathrm{CP-odd}} = \begin{pmatrix}
\dfrac{\mu v_h^2\cos^2\beta}{4v_\Delta}& -\dfrac{v_h\mu\cos^2\beta}{\sqrt{2}} & \dfrac{v_h\mu\cos\beta\sin\beta}{\sqrt{2}} \\
-\dfrac{v_h\mu\cos^2\beta}{\sqrt{2}} & 2 v_\Delta\mu\cos^2\beta & -v_\Delta\mu\sin(2\beta)\\
 \dfrac{v_h\mu\cos\beta\sin\beta}{\sqrt{2}}& -v_\Delta\mu\sin(2\beta) & m_{A^0}^2
\end{pmatrix}
\end{align}
\begin{align}
M_{\mathrm{CP-even}} = \begin{pmatrix}
\dfrac{\mu v_h^2\cos^2\beta}{4v_\Delta}& -\dfrac{\sqrt{3}}{2}\mu v_h\cos\beta& 0 \\
-\dfrac{\sqrt{3}}{2}\mu v_h\cos\beta & m_{h^0}^2 & m_{12}^2\\
0 & m_{12}^2 & m_{H^0}^2
\end{pmatrix}
\end{align}
where 
\begin{align}
    m_{h_0} &= \lambda_1\cos^2\beta + \lambda_6 \sin 2\beta - \frac{m_{22}}{v^2}\tan^2\beta \\
    &+ \sin^2\beta \left(\frac{\lambda_3+\lambda_4+\lambda_5}{2} + \lambda_7\tan\beta + \frac{\lambda_2}{2}\tan^2\beta\right)\nonumber\\
    m_{H_0} &= -\frac{m_{22}}{v^2}+ \frac{\lambda_3+\lambda_4+\lambda_5}{2}\cos^2\beta + \frac{3}{2}\lambda_7\sin2\beta + \frac{3}{2}\lambda_2\sin^2\beta\\
    m_{12} &= \lambda_6\cos^2\beta +\frac{\lambda_3+\lambda_4+\lambda_5}{4} \sin2\beta + \left(\frac{m_{22}}{v^2} - \frac{\lambda_2}{2}\sin^2\beta\right)\tan\beta 
\end{align}
Both Higgs doublets couple to the top quark similarly as in the 2HDM shown in App.~\ref{App: 2hdm}. The RGEs for the quartic and Yukawa couplings are also the same as in this model. 

Defining the mixing angles $\theta$ and $\gamma$ such as the $125\GeV$ Higgs mass eigenstate is defined by
\begin{align}
h^0 &= h_1^0 \sin\gamma\cos\theta + h_2^0\sin\gamma\sin\theta + H_1^0\cos\gamma,
\end{align}
the values of the top Yukawa couplings $y$ and $y'$ at the top quark scale are
\begin{align}
    y &= -\dfrac{y_{SM}}{\sin\theta}\left(\kappa_t\dfrac{\sin\beta}{\sin\gamma} - \dfrac{\sin\theta}{\cos\alpha}\right)\\ 
    y' &= \dfrac{y_{SM}}{\sin\theta}\left(\kappa_t\dfrac{\cos\beta}{\sin\gamma} - \dfrac{\cos\theta}{\cos\alpha}\right)
\end{align}
We can also express $\mu$ in function of the mixing angles $\alpha$, $\beta$, $\gamma$ and $\theta$
\begin{align}
\mu = \dfrac{m_h^2}{v\cos\alpha\cos\beta}\dfrac{2}{\sqrt{2}\cot\alpha\cos\beta-\sqrt{3}\tan\gamma\cos\theta}
\end{align}
The parameters for the benchmark model mentioned in Sec.~\ref{Sec: GM} are
\begin{align}
\beta &= 0.20 \quad \theta = 1.16 \quad \gamma = 1.12 \quad \alpha = 0.52\\
\lambda_1 &= 1.12 \quad \lambda_2 =  0.27\quad \lambda_3 = 2.33\quad \lambda_4 = -1.01 \quad \lambda_5 = -1.01 \quad \lambda_6 = 0 \quad \lambda_7 = -1.60\nonumber
\end{align}

\section{RGEs for yukawa couplings with one top partner and one scalar singlet}
\label{Sec: RGE1top}

The RGEs for a Lagrangian of the type
\begin{align}
    \LL_{\mathrm{int}} = y_0 Hq_3u^c + y_2 Hq_3 U^c  + y_1 S U u^c + M_1 U U^c
\end{align}
where $S$ is taken to be a complex scalar, are 
\begin{align}
    16\pi^2\frac{dy_0}{d\log\mu} &= y_0\left[\frac{9}{2} (y_0^2+y_2^2) + \frac{1}{2} y_1^2 - 8g_3^2\right]\\
    16\pi^2\frac{dy_2}{d\log\mu} &= y_2\left[\frac{9}{2} (y_0^2+y_2^2)+ \frac{1}{2} y_1^2 - 8g_3^2\right] \\
    16\pi^2\frac{dy_1}{d\log\mu} &= y_1\left[4 y_1^2 + y_0^2+y_2^2 - 8g_3^2\right]\\
    16\pi^2 \frac{dg_3}{d\log\mu} &= -\frac{19}{3} g_3^3
\end{align}
therefore
\begin{align}
    y_2(t) &= \frac{y_{2}(0)}{y_{0}(0)}y_0(t)\\
\end{align}
The RGEs would then become
\begin{align}
    16\pi^2\frac{dy_0}{d\log\mu} &= y_0\left[\frac{9}{2} y_0^2 (1+\mathcal{K}^2) + \frac{1}{4} y_1^2 - 8g_3^2\right]\\
    16\pi^2\frac{dy_1}{d\log\mu} &= y_1\left[4 y_1^2 + (1+\mathcal{K}^2)y_0^2 - 8g_3^2\right]\\
    16\pi^2 \frac{dg_3}{d\log\mu} &= -\frac{19}{3} g_3^3
\end{align}
where
\begin{align}
    \mathcal{K} = \frac{y_2(0)}{y_0(0)}
\end{align}
which gives
\begin{align}
    16\pi^2\frac{dy}{d\log\mu} &\gsim \frac{9}{2}(\mathcal{K}^2+1) y_0^3\\
16\pi^2\frac{d|y_1|}{d\log\mu} &\gsim 4 |y_1|^3
\end{align}
so in order to have Landau poles appear only above $10\TeV$, we need
\begin{align}
    y_0(0) &\lsim \frac{2.1}{\sqrt{1+\mathcal{K}^2}}\\
    y_1(0) &\lsim 2.2
\end{align}
Eq.~\ref{Eq: mT} gives
\begin{align}
    \mathcal{K}^2 &= \left(1-\frac{y_{SM}^2}{y_0^2(0)}\right)\left(\frac{m_T^2 y_{SM}^2}{m_t^2 y_0^2(0)}-1\right)
\end{align}
which leaves us with
\begin{align}
    \left(y_0^2(0) - y_{SM}^2\right)\left(\frac{m_T^2 y_{SM}^2}{m_t^2 y_0^2(0)}-1\right) + y_0^2(0)\lsim 2.1^2
\end{align}
\section{RGEs for yukawa couplings with two additional top partners}
\label{Sec: RGE2top}

The RGEs for a Lagrangian of the type
\begin{align}
    \LL_{\mathrm{int}} = y_0 Hq_3u^c + y_1 Hq_3 U^c + y_2 H Q U^c + y_3 H Q u^c + M_1 U U^c + M_2 Q Q^c
\end{align}
are 
\begin{align}
    16\pi^2\frac{dy_0}{d\log\mu} &= y_0\left[\frac{9}{2} (y_0^2 + y_1^2 + y_3^2) + 3 y_2^2 - 8g_3^2\right] + \frac{3}{2}y_1 y_2 y_3\\
    16\pi^2\frac{dy_1}{d\log\mu} &= y_1\left[\frac{9}{2} (y_0^2 + y_1^2 + y_2^2) + 3 y_3^2 - 8g_3^2\right] +  \frac{3}{2}y_0 y_2 y_3\\
    16\pi^2\frac{dy_2}{d\log\mu} &= y_2\left[\frac{9}{2} (y_1^2 + y_2^2 + y_3^2) + 3 y_0^2 - 8g_3^2\right] + \frac{3}{2} y_0 y_1 y_3\\
    16\pi^2\frac{dy_3}{d\log\mu} &= y_3\left[\frac{9}{2} (y_0^2 + y_2^2 + y_3^2) + 3 y_1^2 - 8g_3^2\right] +  \frac{3}{2}y_0 y_1 y_2\\
    16\pi^2 \frac{dg_3}{d\log\mu} &= -\frac{17}{3} g_3^3
\end{align}
These RGEs can also be expressed as
\begin{align}
8\pi^2\frac{d(y_0^2 - y_2^2)}{d\log\mu} &= (y_0^2 - y_2^2)\left[\frac{9}{2}(y_0^2+y_1^2 + y_2^2 + y_3^2) - 8 g_3^2\right]\\
8\pi^2\frac{d(y_1^2 - y_3^2)}{d\log\mu} &= (y_1^2 - y_3^2)\left[\frac{9}{2}(y_0^2+y_1^2 + y_2^2 + y_3^2) - 8 g_3^2\right]
\end{align}
which gives
\begin{align}
    y_0^2(t) - y_2^2(t) &= \mathcal{K}^2 (y_1^2(t) - y_3^2(t))
\end{align}
where
\begin{align}
    \mathcal{K}^2 &= \frac{y_0^2(0) - y_2^2(0)}{y_1^2(0) - y_3^2(0)} \quad t = \log\mu
\end{align}
Therefore, for a given set of initial conditions $\left\{y_{00},y_{10},y_{20},y_{30}\right\}$, the configuration for which Landau poles will appear the latest is always
\begin{align}
|y_{00}| = |y_{01}| = |y_{02}| = |y_{03}|
\end{align}
In this last case, the couplings remain equal at all scales and are therefore described by only one RGE
\begin{align}
16\pi^2\frac{dy}{d\log\mu} &= 15 y^3 - 8yg_3^2
\end{align}
so Landau poles appear later than $10\TeV$ if and only if
\begin{align}
y \lsim 1.06
\end{align}

\bibliography{darkside}
\end{document}